\documentclass[11pt]{iopart}

\usepackage{graphicx}% Include figure files
\usepackage{bm}% bold math
\usepackage{amssymb}
\usepackage{hyperref}
\hypersetup{
colorlinks=true,
linkcolor=black,
citecolor=black,
filecolor=black,
urlcolor=black,
linkbordercolor={1 1 1},
citebordercolor={0 0 0}
}
\newcommand{\theq}{\theta_{eq}}
\newcommand{\vct}[1]{\mathbf{#1}}
\newcommand{\grad}{\bm{\nabla}}
\newcommand{\eqref}[1]{(\ref{#1})}
\newcommand{\sign}{\mbox{sign}}

\begin{document}

\title{Dynamics of nanodroplets on topographically structured
substrates}

\author{A.  Moosavi}
\address{Department of Mechanical Engineering, Sharif University of Technology, Azadi Ave., P.O.Box 11365-9567 Tehran, Iran}
\author{M. Rauscher and S.  Dietrich} 
\address{Max-Planck-Institut f\"ur Metallforschung, Heisenbergstr. 3,
D-70569 Stuttgart, Germany, and} 
\address{Institut f\"ur Theoretische und Angewandte Physik,
Universit\"at Stuttgart, Pfaffenwaldring 57, D-70569
Stuttgart, Germany}
\ead{dietrich@mf.mpg.de}

\date{\today}

\begin{abstract} 
Mesoscopic hydrodynamic equations are solved to investigate the
dynamics of nanodroplets positioned near a topographic step 
of the supporting substrate. Our results show that the dynamics
depends on the characteristic length scales of the system given by the height
of the step and the size of the nanodroplets as well as on the
constituting
substances of both the nanodroplets and the substrate. The lateral motion of
nanodroplets far from the step can be described well in terms of a power law of
the distance from the step. In general the direction of the motion
depends on the details of the effective laterally varying
intermolecular forces.
But for nanodroplets positioned far from the step it is solely
given by the sign of the Hamaker constant of the system.  Moreover,
our study reveals that the steps always act as a barrier
for transporting liquid droplets from one side of the step to the
other. 
%independent of the dynamical status of the droplets near the step.
%whether 
%the droplets near the step are attracted or repelled from the step. 
%We also investigate configuration of nanodroplets positioned on topographic steps. 

\end{abstract}


\submitto{\JPCM}

\maketitle

\tableofcontents

\section{Introduction}

Understanding the wetting behavior of liquids on solid
substrates \cite{degennes85,dietrich88} is a prerequisite for
making use of a myriad  of
biological and technological applications such as eye
irrigation, cell adhesion, tertiary oil recovery, coating,
lubrication, paper industry, micro-mechanical devices, and the
production of integrated circuits. Generically, the solid surfaces in
the above mentioned examples are not ideal in the sense that
they are neither smooth nor homogeneous. Most surfaces are
topographically or chemically heterogeneous.  Such
heterogeneities may substantially change the wetting behavior
of these surfaces \cite{dietrich99}, which is not necessarily
detrimental with respect to envisaged applications. Certain
topographically structured surfaces  are superhydrophobic or
superhydrophilic. In the
first case droplets roll off these substrates (instead of
flowing), such that these surfaces are self-cleaning
\cite{swain98,richard99,lafuma03,krupenkin04,quere05,yang06b,sbragaglia07,quere08}\/.
In the second
case the surface topography leads to a complete spreading of
droplets \cite{bico01,mchale04,extrand07}\/. Tailored
topographic surface structures can induce particular dewetting
processes which in turn can be exploited to pattern substrates on the
micron scale \cite{rockford99,kargupta02c}\/.

Microfluidics is another strong driving force for the research on 
the dynamics of fluids on structured substrates. Shrinking 
standard laboratory setups to a lab-on-a-chip promises 
huge cost reduction and  speed-up
\cite{mitchell01,thorsen02}\/. Open microfluidic systems, i.e., with free
liquid-vapor or liquid-liquid interfaces,
may provide various advantages such as reduced
friction, better accessibility of the reactants, and reduced
risk of clogging by solute particles \cite{zhao01,lam02,zhao02,zhao03}\/.
% \cite{mechkov08,seemann05}. 
%, instead of rigid liquid-solid interfaces, which reduces the frictional forces but 
In open microfluidic devices fluids are guided along chemical
channels \cite{dietrich99,koplik06a,mechkov08} or in grooves \cite{seemann05},
which can be chemically patterned in oder to provide
additional functionality \cite{zhao03}\/.

%the substrates. In contrast to chemically patterning substrates,
%liquids on topographically patterning substrates exhibit a rich
%variety of wetting morphologies and can spread spontaneously along
%these substrates \cite{baret07,khare07,seemann05,brinkmann04b}. 
%%These can make topographically patterning substrates potentially attractive for constructing open microfluidic systems. 
%While present devices are based mostly on micron sized channels,
%further miniaturization will eventually lead to devices in
%nanoscales. Since there are many fundamental differences between
%the behavior of fluids on the nanoscales with those on the
%microscales or above  \cite{karniadakis05,dietrich05} it is
%critical to understand basic fluidic issues occurring at those
%scales.  Recent theoretical studies of nanoscale fluids on
%chemically \cite{cieplak06,moosavi08a,moosavi08b} and
%topographically \cite{moosavi06b} structured substrates have
%underscored the importance of such investigations.

Wetting phenomena on topographically structured substrates
have attracted substantial research efforts
\cite{seemann05,robbins91,netz97,rejmer99,rascon00,bruschi02,klier05,gang05,ondarcuhu05,tasinkevych06a,tasinkevych07b}
with, however, the main focus on equilibrium phenomena.
In view of the aforementioned applications, dynamical aspects are of particular
interest. 
%Since the performance of these systems ultimately depend on their
%dynamic properties, the investigations concerning the dynamics are
%of particular interest.
%an important prerequisite 
In spite of this demand, theoretical work on the dynamics of
liquid films and droplets on topographically structured substrates
has started only recently.
In most of these studies the dynamics of the fluids is assumed to be well
described by macroscopic hydrodynamic equations, which are solved either
directly \cite{gramlich04}, by a lattice Boltzmann method 
\cite{dupuis04,dupuis05b}, or in the thin film (lubrication) regime
\cite{davis05,gaskell06,saprykin07,tseluiko08}\/.
The applicability of this latter method is limited 
because the inherent long-wavelength approximation does not keep
track of many relevant microscopic features \cite{oron00a}\/. 

On the nanoscale, macroscopic hydrodynamic equations turn out to
be inadequate for describing the dynamics of fluids. 
Overcoming this deficit is the focus of a new
research area called nanofluidics \cite{eijkel05,mukhopadhyay06}\/.
Wetting phenomena in particular reveal these
deviations; for a recent review of these issues see Ref.~\cite{rauscher08a}\/.
However, hydrodynamic equations can be augmented to include 
hydrodynamic slip, the finite range of intermolecular
interactions, and thermal fluctuations. The resulting mesoscopic
hydrodynamic equations have been rather successful in analyzing, e.g., 
the dynamics of dewetting on homogeneous substrates
\cite{fetzer06a,fetzer07b}\/. The presence of 
intermolecular interactions can be summarized into the so-called
disjoining pressure (DJP), $\Pi = -\partial\Phi/\partial y$ where
the effective interface potential $\Phi$ is the cost of free energy
to maintain a homogeneous wetting film of prescribed thickness $y$\/.
On a homogeneous substrate $\Phi$ is independent of lateral
coordinates parallel to the substrate surface and the equilibrium
wetting film thickness $y_0$ minimizes $\Phi(y)$\/.
However, on chemically or topographically
inhomogeneous substrates (structured, rough, or dirty) the
generalized disjoining pressure does depend in addition on these lateral
coordinates. In most studies, the lateral variations
of the disjoining pressure have been modelled rather crudely, i.e., the
substrate is assumed to be locally homogeneous and lateral
interferences of heterogeneities are neglected: e.g., a step is
typically modelled by an abrupt change of the disjoining pressure
\cite{brusch02,bielarz03,thiele03a,gaskell04a,yochelis07}\/. 

\begin{figure}
\includegraphics[width=0.8\linewidth]{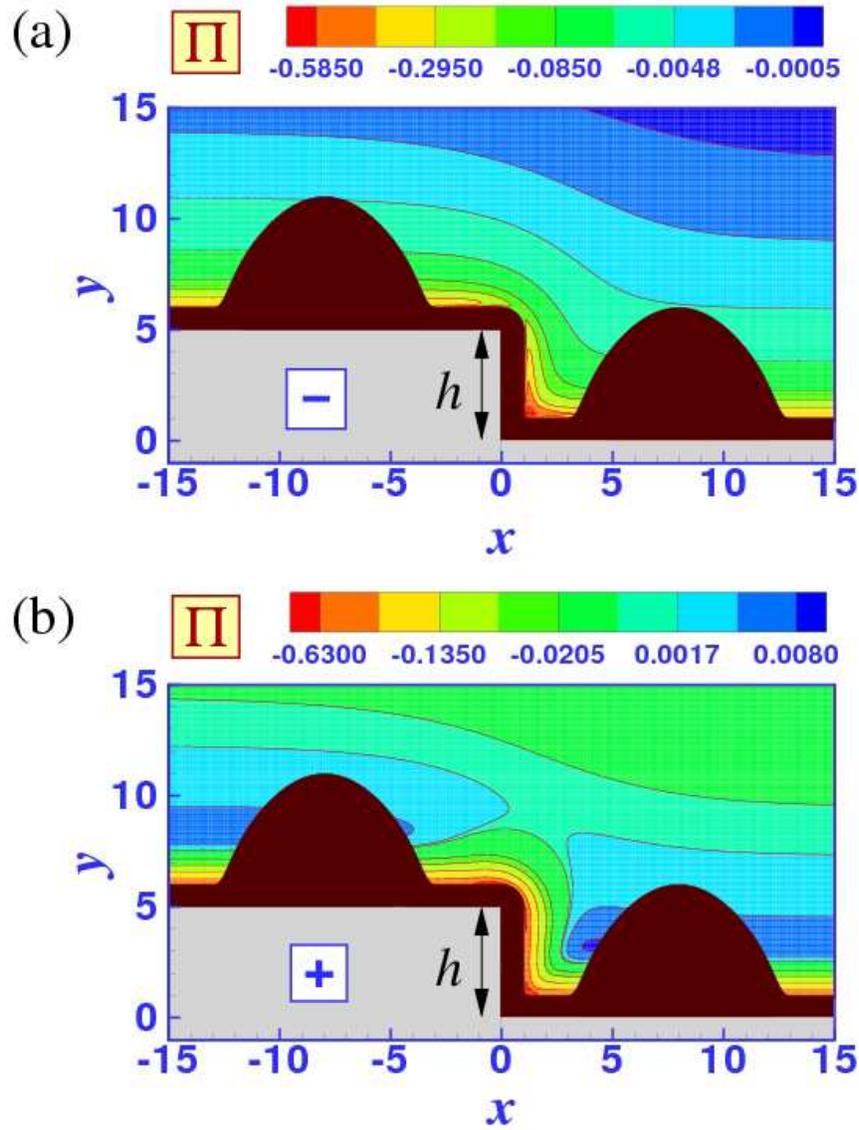}
\caption{Nanodroplets positioned near a topographic step of height
$h$ (on the top side and bottom side of the step) are exposed to the vertically 
and laterally varying disjoining pressure $\Pi$, the contour plot
of which is shown.  The topographic step and the drops are taken to
be translationally invariant along the $z$ axis (i.e., orthogonal
to the image plane). In (a) the substrate is chosen to correspond
to the minus case \fbox{$-$}
with ($B=0$, $C=1$) and in (b) the substrate corresponds to the plus 
case \fbox{$+$} ($B=-2.5$, $C=1$)
(see Eq.~\protect\eqref{djpfar} for definitions)\/.
Lengths ($x$, $y$, $h$) and the disjoining pressure $\Pi$ are measured in units of $b$ 
and $\gamma/b$, respectively (see the main text for definitions). }
\label{dpstep}
\end{figure}

Recently we have demonstrated, that the actually smooth variation
of the lateral action of
surface heterogeneities can change the behavior of droplets in the
vicinity of chemical steps \cite{moosavi08a,moosavi08b} or
topographical features (edges and wedges) \cite{moosavi06b} even
qualitatively.
In the present study we extend these results to the case of an
isolated straight topographic step in an otherwise homogeneous
substrate (as shown in Fig.~\ref{dpstep}) 
and we recover the previously studied case of isolated wedges and
edges in the limit of infinite step height $h$\/. 
%above mentioned phenomena \cite{moosavi06b} by developing the
%results obtained for edges and wedges to nanometric steps. Our
%results in the limiting cases, that is, when the height of the step
%approaches to infinity, resembles those of edges and wedges
%\cite{moosavi06b}. 
We should emphasize that our investigation
provides only a first but nonetheless essential step towards
understanding the dynamics of droplets on arbitrarily structured
substrates. Although more refined than previously used models
the present one is still rather simple. We only
consider additive Lennard-Jones type intermolecular interactions,
i.e., we do not take into account electrostatic interactions which
would be very important for polar fluids. We assume the fluid to be
Newtonian, non-volatile, and incompressible (which is compatible
with the frequently used so-called sharp-kink approximation of classical equilibrium density
functional theory (see, e.g., Ref.~\cite{bauer99a})\/. We also assume a
no-slip boundary condition at the solid surface \cite{lauga05a} and
neglect the influence of thermal fluctuations \cite{gruen06a}\/.
For numerical reasons we
restrict our investigation to two-dimensional (2D)
droplets, corresponding to three-dimensional (3D) liquid ridges (or
rivulets) which are translationally invariant in the direction
parallel to the step; nonetheless we expect our results to hold
qualitatively also for 3D droplets.
%(such liquid ridges have been experimentally
%studied, e.g., in Ref.~\cite{ondarcuhu91})\/.  It is expected that
%our results will carry over qualitatively to the behavior of 3D
%droplets.  We propose these issues for further studies. 

%The paper is organized as follows: The mesoscopic hydrodynamic
%equations underlying our study are developed in the next section
%(Sec.~\ref{mesohydrosec}) and details of the modelling of the DJP
%in the vicinity of topographic steps are presented in
%Sec.~\ref{Modeling}\/. 
%The results of the numerical solution of the mesoscopic
%hydrodynamic equations are presented in
%Sec.~\ref{results} and discussed in terms of the forces acting on
%nanodroplets in Sec.~\ref{discuss}\/. 
%Finally Sec.~\ref{Summary} contains a summary of our study and an
%outlook on perspectives of this research area.
%The numerical boundary element algorithm used for
%investigating the dynamics is presented in detail in
%\ref{numersec}\/.

%==========================   FIGURE  ================================== 
%======================================================================= 

\section{Summary}
We study the dynamics of non-volatile and
Newtonian nanodroplets (corresponding to
three-dimensional ridges which are translationally
invariant in one lateral direction) on topographically stepped
surfaces within the framework of
mesoscopic hydrodynamics, i.e., by solving the augmented Stokes
equation presented in Sec.~\ref{mesohydrosec} with the numerical
method described in \ref{numersec}\/. 
We consider in particular the effects due to the long range
of Lennard-Jones type 
intermolecular interactions which enter the theoretical description 
in terms of the disjoining pressure
(DJP) as illustrated in Fig.~\ref{dpstep}\/. We assume the
substrate to be chemically homogeneous in the lateral directions
and the surface to be covered by a thin layer of a different
material. As detailded in Sec.~\ref{Modeling} this leads to two adjustable parameters $B$ and $C$ which
enter into the DJP and characterize the wetting
properties of the substrate, i.e., the equilibrium contact angle
$\theq$ and the wetting film thickness $y_0$ (see
Fig.~\ref{djpshapes})\/. As shown in Fig.~\ref{bc} both 
for positive and for
negative Hamaker constants one can find a one-parameter family of
pairs $(B,C)$ leading to the same $\theq$ on a flat
substrate (i.e., without a step)\/. As shown in Fig.~\ref{flat} 
nanodroplets on
substrates with the same $\theq$ but with different values of $B$
and $C$ assume shapes which 
differ mainly in the vicinity of the three-phase contact
line with the apex region is almost unaffected by the substrate
potential.

The results of the numerical solution of the mesoscopic
hydrodynamic equations are presented in
Sec.~\ref{results}\/.
In contrast to macroscopic expectations
based on a capillary model (i.e., taking into account only 
interface energies and neglecting the long range of the
intermolecular interactions), topographic steps do influence droplets
in their vicinity: on substrates with a positive Hamaker constant
(Figs.~\ref{effstepplus} and \ref{wedgediffhplus}),
droplets move in uphill direction while on substrates with a
negative Hamaker constant (Figs.~\ref{effstepp} and
\ref{wedgediffhminus}) the droplets move in the opposite direction.
As expected the forces on the droplets and their resulting velocity
increase with the step height, but also with the absolute value of
the Hamaker constant. This is the case if the contact angle is
varied (see, e.g., Figs.~\ref{edgeeffc} and \ref{wedgediffcminus})
and even if the contact angle is fixed by varying the Hamaker
constant and the properties of the coating layer together
(see Figs.~\ref{edgediffb}, \ref{edgeplus}, \ref{wedgediffb}, and
\ref{wedgediffbplus})\/. 

The speed of the droplets increases with their size as
demonstrated in Fig.~\ref{wedgedropsize}\/. As detailed in
Subsec.~\ref{direction}, the influence of the step on a droplet can be
phrased in terms of an effective wettability gradient, i.e., a spatially
varying equilibrium contact angle. The driving force on droplets on
such substrates increases linearly with the droplet size because the
difference in equilibrium contact angle at the two contact lines
of the liquid ridges
increases roughly linearly with the distance from the steps. 

The velocity of droplets driven away from the step decreases
rapidly with the distance from the step as shown in
Sec.~\ref{discuss}\/. But droplets moving towards the step
(either on the top side or on the bottom side of the step)
stop with their leading contact
line close to the step edge or wedge, respectively. Therefore they do not
cross the step (see Figs.~\ref{effstepp}, \ref{comppisigma},
\ref{edgeeffc}, \ref{edgediffb}, \ref{wedgediffhplus}, and
\ref{wedgediffbplus})\/. Accordingly,
edges, wedges, and steps act as barriers for migrating droplets
(which is also true macroscopically) because droplets sitting right
at the tip of an edge are in an free-energetically unfavorable state
(see Fig.~\ref{overedge}) while droplets located in the corner of 
a wedge are in a state corresponding to a 
local minimum of the free energy (see Fig.~\ref{wedgeforce})\/.
Therefore, an external force is required to push droplets
over edges (see Fig.~\ref{edgeforce}) or to pull them out of wedges
(see Fig.~\ref{wedgeforce})\/. In both cases, the total (i.e.,
integrated over the droplet volume) force required to accomplish
this increases slightly with the droplet volume, but less
than linearly. This means, that if the force is applied via a body
force density acting per unit volume (e.g., gravity) larger droplets
experience a larger force and therefore overcome steps more easily. 
In addition, the lateral action of
intermolecular forces can also pin droplets at edges and near
wedges. However, droplets which initially span a topographic step
always end up filling the wedge at the step base, either with the
upper contact line pinned at the step edge or, if the droplet
volume is too small, with the upper contact line on the vertical
part of the step, as shown in Fig.~\ref{overwedge}\/. 

A deeper understanding of the dynamics of droplets in the vicinity
of edges and wedges can be reached by analyzing the forces acting
on the droplet surface, i.e., the disjoining pressure and surface tension
(see Eqs.~\eqref{dpforce} and \eqref{sigmaforce}, respectively)\/. As
demonstrated in Fig.~\ref{comppisigma}, if the droplets move
under the influence of the topographic step only, the main
contribution to the driving
force stems from the disjoining pressure. As shown in Figs.~\ref{dropsize}
and \ref{stepsize}, the numerically observed features of the
dynamics of droplets can be understood in terms of the disjoining
pressure induced force density on the droplets calculated for droplets of
simple parabolic shapes used as initial conditions for the numerical
solution of the hydrodynamic equations. As shown in
Fig.~\ref{profsize} the actual relaxed droplet
shape is different but the calculated forces depend only weakly 
on the deviation of the actual shape from its parabolic
approximation. 
In the
limit of large distances from the step the force can be calculated
analytically (see Subsec.~\ref{direction}):
far from the step the total force per unit ridge length
$F_\Pi=f_\Pi\,A_d$ (with the cross-sectional area $A_d$)
essentially depends on the ratio of the
step height $h$ and the distance from the step $\bar{x}$ as well
as on the ratio of the apex height $y_m$ and $\bar{x}$\/. The
corresponding asymptotic
results are summarized in Fig.~\ref{asymptofig}\/. In all cases the
force density varies according to a power law $\bar{x}^{-\zeta}$
with $\zeta \in \{3,4,5\}$\/.
For finite sized droplet and steps of finite height 
we obtain the fastest decay and
for almost macroscopic droplets
in the vicinity of finite sized steps as well as for nanodroplets
near isolated edges and wedges we get $\zeta=4$\/. While our present
analysis cannot be applied to the case of an almost macroscopic
droplet in a wedge, for large drops ($y_m/\bar{x}\to \infty$)
next to an isolated edge we get
the weakest decay with $\zeta=3$\/. In any case, the 
total force per unit length $F_\Pi$ is proportional to the Hamaker
constant as observed in the numerical solution of the mesoscopic
Stokes dynamics as well as in the force analysis presented 
in Subec.~\ref{force}\/. 
The dynamics of large drops ($y_m/\bar{x}\to \infty$) 
is equivalent to the dynamics of macroscopic drops
on a surface with an effective chemical wettability gradient (i.e., a
spatially varying ``equilibrium contact angle'' $\theq(x)$)
\cite{chaudhury92,subramanian05,pismen06}\/.

\section{Mesoscopic hydrodynamic equations}
\label{mesohydrosec}

At low Reynolds numbers the mean field dynamics of an incompressible
Newtonian fluid of viscosity $\mu$ is given by the
Navier-Stokes equation for the local pressure $p(\vct{r},t)$ and the
flow field $\vct{u}(\vct{r},t)$:
\begin{eqnarray}
\label{eq:stokes}
\grad\cdot\bm{\sigma}& =& -\grad p+  \mu \,\grad^2 \vct{u}=0,\\
\label{eq:incomp}
\grad\cdot \vct{u}&=&0,
\end{eqnarray}
with the stress tensor $\sigma_{ij}=-p\,\delta_{ij}+
\mu\,(\partial_j u_i+\partial_i u_j)$\/.
%At the surface of the impermeable substrate
%we assume a no-slip boundary condition 
%\begin{equation}
%\vct{u}=0.
%\end{equation}
In this study, we neglect the influence of the vapor phase or air
above the film. Therefore the tangential components of the
component of the stress tensor $\bm{\sigma}\cdot\vct{n}$ 
normal to the liquid-vapor surface
$\Gamma_{lv}$ (with outward pointing normal vector
$\vct{n}$) is zero. The normal component of
$\bm{\sigma}\cdot\vct{n}$, i.e., the normal forces acting on the
liquid surface, are given by the sum of the Laplace pressure and
of the disjoining pressure:
\begin{equation}
\label{eq:surfacebc}
\bm{\sigma}\cdot\vct{n} =
\vct{n}\,\left(\gamma\,\kappa+\Pi+x\,g\right)
\quad\mbox{at}\quad\Gamma_{lv},
\end{equation}
with the surface tension coefficient $\gamma$ and the local mean
curvature $\kappa$  of the liquid surface; 
$g$ is the strength of a spatially constant external body force
density pointing in the $x$-direction
(with $-g\,x$ as the corresponding potential) 
which we introduce in order to study the strength of
barriers to the lateral motion of droplets. Alternatively,
for incompressible fluids one can define a new pressure
$p'=p-x\,g$ such that the external body force density $g$ enters
into the Stokes
equation Eq.~\eqref{eq:stokes} rather than the boundary condition
in Eq.~\eqref{eq:surfacebc}: $0=-\grad p'+\vct{e}_x\,g +
\mu\grad^2\vct{u}$ \/. Although this approach might be more
intuitive, the equivalent form used here is more convenient for
implementing the boundary
element method used here to numerically solve these equations (see
\ref{numersec})\/.

The dynamics of the free liquid surface is determined by mass
conservation together with the incompressibility condition: the
local normal velocity is identical to the normal component of the
local flow field.

We neglect hydrodynamic slip at the liquid-substrate surface $\Gamma_{ls}$ and we only consider impermeable
substrates. Since we assume the substrate to be stationary this
results in the following boundary condition for the flow field:
\begin{equation}
\vct{u}=0 \quad\mbox{at}\quad \Gamma_{ls}.
\end{equation}

In order to avoid strong initial 
shape relaxation of the droplets (in response to placing them on the
substrate with a certain shape) which can lead to significant lateral 
displacements \cite{ondarcuhu92}, we choose a parabolic initial profile
which is smoothly connected to a precursor film of thickness $y_0$:
\begin{equation}
\label{inicond}
y(x;t=0)= y_0+a\,\left[1-\left(\frac{|x-\bar{x}|}{a}\right)^2\,
\right]^{|x-\bar{x}|^m+1},
\end{equation} 
such that $a$ is the droplet height at the center and half the base width. 
Accordingly the distance of the droplet edge from the step at $x=0$ 
is given by 
$\ell=|\bar{x}|-a$  with $\bar{x}$ the position of the center 
of the droplet in the $x$-direction. % (see Fig.~\ref{edgediffb})\/. 
The parameter $m$ specifies the smoothness of the transition region 
from the drop to the wetting layer. In this study we choose $m$ to be $10$\/. 
We investigate the droplet dynamics for two different situations. 
In the first one we position the droplet on the top side of the
step of height $h$ with the 
three-phase contact line $(x=\bar{x}+a,\,y=h+y_0,z)$ at a distance 
$\ell=-\bar{x}-a$ with $\bar{x}<-a$ from the step edge at $x=0$\/.
In the second situation we place 
the droplet on the bottom side of the step with the three-phase contact 
line $(x=\bar{x}-a,\,y=y_0,z)$ at a distance $\ell=\bar{x}-a$ with
$\bar{x}>a$ from the wedge at the base of the step. 

In equilibrium Eq.~\eqref{eq:surfacebc} reduces to the
Euler-Lagrange equation of the effective interface Hamiltonian of a 
fluid film on a substrate as derived, e.g., in
Ref.~\cite{dietrich91b}\/. This means that we approximate the normal
forces on the liquid surface due to the intermolecular interactions by the
disjoining pressure derived for equilibrium systems.

In a non-equilibrium situation, the unbalanced forces acting on the fluid
surface add up to a resulting net force on the liquid body. We
separately consider the two
contributions $f_{\Pi}$ and $f_{\gamma}$ from the disjoining
pressure and from the Laplace pressure, respectively, both
normalized by the droplet volume $\Omega_d$ and given by the
following integrals over the liquid-vapor surface $\Gamma_d$ of the
droplets:
\begin{eqnarray}
\label{dpforce}
f_{\Pi}(x)&=&\frac{1}{\Omega_d}\int_{\Gamma_d}\Pi(x,y)\,n_{x}ds\\
\label{sigmaforce}
f_{\gamma}(x)&=&\frac{1}{\Omega_d}\int_{\Gamma_d}\gamma\,\kappa\,n_{x}ds\,.
\end{eqnarray}
%In the case of a liquid ridge translationally invariant in
%$z$-direction these expressions yield a force per unit ridge
%length.
For a liquid ridge translationally invariant in
$z$-direction both integrals as well as $\Omega_d = A_d\,L$ are proportional to
the macroscopic ridge length $L$, so that the latter drops out of the
expressions for the force densities (in units of
$\mbox{N}/\mbox{m}^3$) $f_{\Pi}$ and $f_{\gamma}$\/. In three dimensions
$ds$ is a two-dimensional surface area element. $A_d$ is the
two-dimensional cross-sectional area of the liquid ridge.

\section{Model of the heterogeneity}
\label{Modeling}
In the following we calculate the disjoining pressure for a fluid
film or droplet near a topographic step as displayed in 
Fig.~\ref{dpstep}\/. Apart from a very thin coating layer of
thickness $d$ we assume the substrate material to be homogeneous,
disregarding its discrete molecular structure.
Many substrates used in experiments are coated, e.g., by a native
oxide layer or by a polymer brush which is used to modify the
wetting properties of the substrate. 
However, a more refined analysis of the DJP, which
takes the molecular structure of the substrate and of the fluid into
account, yields terms of a form similar to those generated
by a coating layer \cite{dietrich88,dietrich91a}\/.
In general, i.e., far from the critical point of the fluid, 
the vapor or gas phase covering the system has a negligible density
which we neglect completely.
Assuming pairwise additivity of the intermolecular interactions, i.e., 
the fluid particles as well as the fluid and the 
substrate particles are taken to interact with each other via
pair potentials $V_{\alpha\beta}(r)$ where $\alpha$ and $\beta$ 
relate to liquid ($l$), substrate ($s$), or coating ($c$) 
particles and $r$ is the interatomic distance, one can show 
that the disjoining pressure (DJP) of the system is given by \cite{robbins91}
\begin{equation}
\label{djpg}
\Pi({\bf {r}})=
\int_{\Omega_s}
{\left[\rho_l^2\,V_{ll}({\bf{r}}-{\bf{r'}})
-\rho_l\,\rho_s\,V_{sl}({\bf{r}}-{\bf{r'}})\right]}\,
\,d^3r,
\end{equation} 
with $\bf{r}, \bf{r'} \in \mathbb{R}^{\textnormal{3}}$ and 
$\rho_l$ and $\rho_s$ as the number densities of the liquid 
and substrate, respectively. $\Omega_s$ is the actual substrate volume. 

In order to facilitate the calculation of the disjoining pressure 
of the step we decompose it into contributions from
quarter spaces (edges) forming building blocks which can be
calculated analytically.
We first consider an $e$dge occupying the lower left quarter space 
$\Omega_e^{\lhd}=\lbrace \vct{r} \in \mathbb{R}^3 \mid x\leq 0 \wedge y
\leq0\rbrace $, which in the following we denote by $\lhd$\/.
%that is, the one in the left part of the step that we denote by $\lhd$. 
For Lennard-Jones type pair potentials $V_{\alpha
\beta}(r)={M_{\alpha \beta}}/{r^{12}}-{N_{\alpha
\beta}}/{r^6}$, where $M_{\alpha \beta}$
and $N_{\alpha \beta}$ are material parameters, the DJP in the
vicinity of a non-coated edge occupying $\Omega_e^{\lhd}$ is given
by
\begin{equation}
%\begin{multline}
\label{eqpie}
\Pi^{\lhd}_e(x,y)=\int\limits_{-\infty}^{0}dx'
\int\limits_{-\infty}^0dy'\int\limits_{-\infty}^{-\infty}dz'\left(
\frac{\Delta M_e}{\left| \vct{r}-\vct{r}'\right|^{12}
}-\frac{\Delta N_e}{\left|
\vct{r}-\vct{r}'\right|^6 }\right),
%\end{multline}
\end{equation}
where $\Delta M_e= \rho_l^2\,M_{ll}-\rho_l\,\rho_s\, M_{ls}$ and
$\Delta N_e=\rho_l^2\,N_{ll}-\rho_l\,\rho_s\, N_{ls}$\/.  
The first term dominates close to the surface of the edge and the
second term at large distances from the substrate.

All integrals in Eq.~(\ref{eqpie}) can be calculated 
analytically and one obtains the DJP as the corresponding difference 
$\Pi_e^{\lhd}= \Delta M_e\,I_e^{12\lhd}-\Delta N_e\,I_e^{6\lhd}$ of two contributions with
\begin{eqnarray}
I^{12}_e(x,y)&=&\frac{\pi}{11520\,x^9\,y^9\,(x^2+y^2)^{7/2}}
[-280\,x^6\,y^6(x^4+\nonumber\\
&&y^4)-448\,x^2y^2(x^{12}+y^{12})-128\,(x^{16}+\nonumber\\
&&y^{16})+128\,(x^9+y^9)(x^2+y^2)^{7/2}\nonumber\\
&&-35\,x^8\,y^8-560\,x^4y^4\,(x^8+y^8)]\nonumber\\
\end{eqnarray} 
and 
\begin{eqnarray}
I^{6}_e(x,y)&=&\frac{\pi}{24\,x^3\,y^3 \sqrt{x^2+y^2}}
[2\,(x^3+y^3)\sqrt{x^2+y^2}\nonumber\\
&&-2\,(x^4+y^4)-y^2\,x^2]\cdot
\end{eqnarray} 

The contributions to the disjoining pressure of a thin 
$c$oating layer of thickness $d$ on the $u$pper side of the edge occupying 
$\Omega_c^{u\lhd}=\lbrace {\bf{r}} \in \mathbb{R}^3
\mid x\leq 0,-d\leq y\leq0\rbrace$, the $r$ight part of the edge 
occupying $\Omega_c^{r\lhd}=\lbrace {\bf{r}} \in \mathbb{R}^3\mid
-d\leq\ x\leq 0, y\leq0 \rbrace$, and  the
thin rod which fills the $t$ip area of the edge
$\Omega_c^{t\lhd}=\lbrace {\bf{r}} \in \mathbb{R}^3 \mid
-d\leq x\leq 0,-d\leq y\leq0\rbrace$
can be calculated analogously:
\begin{eqnarray}
\label{picoat}
\Pi^{\chi\lhd}_c(x,y)=\int_{\Omega_c^{\chi\lhd}}
\frac{\Delta M_c}{\left|
\vct{r}-\vct{r}'\right|^{12} }\,d^3r' -\int_{\Omega_c^{\chi\lhd}}
\frac{\Delta N_c}{\left|
\vct{r}-\vct{r}'\right|^6 }\,d^3r', \nonumber \\
\end{eqnarray}
with $\Delta\,M_c=\rho_l^2
M_{ll}-\rho_c\,\rho_l\, M_{cl}$ and $\Delta\,N_c=\rho_l^2\,
N_{ll}-\rho_c\,\rho_l\, N_{cl}$; $\chi$ stands for $u$ ($u$pper), 
$r$ ($r$ight), or $t$ ($t$ip)\/. 
%Similar to the DJP of an edge, the DJP of the coatings can be expressed as $\Pi^{\chi\lhd}_c= \Pi_c^{12\chi\lhd}-\Pi_c^{6\chi\lhd}$. 
Actual coating layers 
have a more complicated structure, in particular in the direct
vicinity of edges and wedges, which 
depends on the specific combination of coating and substrate material as well as
on the way the coating is produced. Such details can influence 
droplets if their contact line is right at the edge or wedge 
but the effect is proportional to the square of the coating layer
thickness $d$\/.  For simplicity we only consider systems with
coating layers which are thin compared to the wetting film
thickness (see below), for which the contribution from the thin rod of coating
material at the tip of the edge or in the corner of the wedge is irrelevant.
According to Eq.~\eqref{picoat} 
the contribution to the disjoining pressure from the upper coating
layer can be decomposed into 
$\Pi_c^u=\Delta M\,I_c^{12u}(x,y)-\Delta N\,I_c^{6u}(x,y)$\/.  To
first order in $d$ we obtain 
%\begin{subequations}
\begin{eqnarray}
I^{12u}_c(x,y)&=&\frac{\pi\,d}{1280\,{(x^2+y^2)}^{9/2}\,y^{10}}\times\nonumber\\
&&[128\,(x^2+y^2)^{9/2}-315\,x\,y^8-840\,x^3\,y^6-\nonumber\\
&&1008\,x^5\,y^4-576\,x^7\,y^2-128\,x^9]
\end{eqnarray}
and
\begin{eqnarray}
I^{6u}_c(x,y)&=&\frac{\pi\,d}{8\,y^4\,{(x^2+y^2)}^{3/2}}\times\nonumber\\
&&[-2\,(x^2+y^2)^{3/2}+3\,x\,y^2+2\,x^3].
\end{eqnarray}
%\end{subequations}
By symmetry one has 
$\Pi^{r\lhd}_c(x,y)=\Pi^{u\lhd}_c(y,x)$
for the contribution of the vertical part of the coating\/. 
The DJP of a coated edge occupying 
$\Omega_{ce}^{\lhd}=\lbrace \vct{r} \in \mathbb{R}^3 \mid x\leq 0 \wedge y
\leq 0 \rbrace $ is therefore given by
\begin{equation}
\Pi^{\lhd}_{ce}(x,y)=\Pi^{\lhd}_e(x+d,y+d)+\Pi^{u\lhd}_c(x,y)+
\Pi^{r\lhd}_c(x,y).
%&&+\Pi_c^{t\lhd}(x,y).
\end{equation}
%%This equation  can be rewritten as 
%%\begin{eqnarray}
%%\Pi^{\lhd}_{ce}(x,y)&=&\Pi^{\lhd}_e(x,y)-\Pi^{u\lhd}_e(x,y)-
%%\Pi^{r\lhd}_e(x,y)\\\nonumber
%%&&+\Pi^{u\lhd}_c(x,y)+\Pi^{r\lhd}_c(x,y).
%%%+\Pi_c^{t\lhd}(x,y),
%%\end{eqnarray}
%%where $\Pi^{u\lhd}_e(x,y)$ and $\Pi^{r\lhd}_e(x,y)$ are the 
%%effect of coating layers of the edge when they are filled 
%%with the material of the edge (in Eq.~\eqref{picoat} replace subscript $c$ by $e$). 

The DJP contribution from a coated edge occupying the right quarter 
space $\Omega^{\rhd}_{ce}=\lbrace \vct{r} \in \mathbb{R}^3 \mid x\geqslant 
0 \wedge y\leq 0 \rbrace $ can be obtained analogously. However, 
since the integrals for the right part 
corresponding to Eqs.~\eqref{eqpie} and \eqref{picoat} are 
the mirror image (with respect to the $yz$-plane) of their counterparts
for the left hand side, the former ones can be expressed in terms of the 
latter ones. Therefore the DJP of the coated lower right quarter space
$\Pi^{\rhd}_{ce}(x,y)$ is equal to $\Pi^{\lhd}_{ce}(-x,y)$\/. 
Combining the contributions of 
the left and the right part leads to the following expression for the 
DJP of a step of height $h$:
\begin{equation}
\label{dpstepeq}
 \Pi(x,y)=\Pi^{\lhd}_{ce}(x,y+h)+\Pi^{\lhd}_{ce}(-x,y)-2\,\Pi^{r\lhd}_c(x,y).
\end{equation}
The last term on the right hand side of 
Eq.~\eqref{dpstepeq} removes the artificial extra coatings on the left and 
the right quarter spaces (at $x=0$, $y<0$) which get buried upon
building the step out of the coated edges. Figure~\ref{dpstep}
shows typical examples for the DJP\/. 
The DJP is not only a function of the vertical distance from the
substrate, but also of the lateral distance from the step. 
In this regard, the substrate in the vicinity of the step resembles a 
chemically structured substrate with laterally varying wettability 
\cite{chaudhury92,subramanian05,pismen06}\/.
%in which the degree of wettability gradients is a function of distance from the step. 

For positions far from the step the distribution of the DJP resembles 
that of the $c$oated, laterally $h$omogeneous flat substrate obtained
by setting $h=0$ in Eq.~\eqref{dpstepeq}\/. To linear order in $d$
one has
\begin{eqnarray}
\Pi_{ch}(y)&=&\frac{\pi\,\Delta
M_e}{45\,y^9}-\frac{\pi\,\Delta N_e}{6\,y^3}-\frac{\pi\,\Delta
M_e\,d}{5\,y^{10}}+\frac{\pi\,\Delta N_e\,d}{2\,y^4}\nonumber\\
&&+\frac{\pi\,\Delta
M_c\,d}{5\,y^{10}}-\frac{\pi\,\Delta N_c\,d}{2\,y^4}.
\end{eqnarray}
Since the repulsive contributions decay rapidly 
with distance from the substrate we neglect all 
those repulsive contributions which are shorter ranged than the corresponding 
term ($\sim y^{-9}$) arising from $\Pi_e^{12\lhd}(x,y)$ 
\cite{bauer99a,dietrich91a}, leading to
\begin{eqnarray}
\label{piflat}
\Pi_{ch}(y)&=&\frac{\pi\,\Delta
M_e}{45\,y^9}-\frac{\pi\,\Delta N_e}{6\,y^3}-
\frac{\pi\,\Delta N\,d}{2\,y^4}\,,
\end{eqnarray}
with $\Delta N=\Delta N_c-\Delta N_e$\/. 
The equilibrium thicknesses $y_0$ of the wetting film on
such a substrate minimizes the effective
interface potential \cite{dietrich91a,napiorkowski92} 
\begin{equation}
\label{effp}
\Phi_{ch}(y)=\int_y^{\infty}\Pi_{ch}(y)dy.
\end{equation}
With Eq.~\eqref{piflat} this leads to
\begin{eqnarray}
\label{eflat}
\Phi_{ch}(y)&=&\frac{\pi\,\Delta
M_e}{360\,y^8}-\frac{\pi\,\Delta
N_e}{12\,y^2}-\frac{\pi \,\Delta N\,d}{6\,y^3}.
\end{eqnarray}
The second term is usually written 
as $-{H_e/(12\,\pi\, y^2)}$, where
$H_e=\pi^2\,\Delta N_e$ is the so-called Hamaker constant.

At this point we introduce dimensionless quantities 
(marked by $*$)  such that lengths
are measured in units of $b={[2|\Delta M_e|/(15\,|\Delta N_e|)]}^{1/6}$ which 
for $\Delta M_e>0$ and $\Delta N_e>0$ is the equilibrium 
wetting film thickness $y_0$ on the uncoated flat substrate. 
The DJP is measured in units of the ratio $\gamma/b$ 
where $\gamma$ is the liquid-vapor surface tension.
Thus the dimensionless DJP $\Pi_{cf}^*=\Pi_{cf}\,b/\gamma$ 
far from the edge has the form 
\begin{equation}
\label{djpfarstar}
\Pi_{ch}^*(y^*)={C}\left(\mp\frac{1}{{y^*}^9} \mp
\frac{1}{{y^*}^3} + \frac{B}{{y^*}^4}\right)\cdot
\end{equation} 
In the first and second term of Eq.~(\ref{djpfarstar}) 
the upper (lower) sign corresponds 
to $\Delta M_e<0$ $(\Delta M_e>0)$ and $\Delta N_e<0$ $(\Delta N_e>0)$, respectively. 
The dimensionless amplitude $C=A\,b/\gamma$, with
$A=\pi{(|\Delta M_e|/45)}^{-1/2}{(|\Delta N_e|/6)}^{3/2}$, compares
the strength of the effective intermolecular forces in the uncoated case 
and of the surface tension forces. The amplitude $B=\pi \Delta
N\,d/(2\,A\,b^4)$ measures the strength of the
coating layer. Since the molecular structure of the substrate and of the fluid 
yields a term of the same form \cite{dietrich88,dietrich91a} we
consider $B$ itself as a parameter independent of the actual properties of the
coating layer. For the 
interactions considered here, $\Delta M_e\ge0$ is a necessary 
condition for the occurrence of an equilibrium wetting layer of 
nonzero thickness but $\Delta N_e$ can be 
positive or negative. Therefore the first term in
Eq.~\eqref{djpfarstar} can only be positive while the second term
can be positive or negative. In the following we shall refer 
to these two cases simply as the minus (\fbox{$-$}) and the plus
(\fbox{$+$}) case. 
In order to avoid a clumsy notation in the following we also drop the stars. 
With this, one has
\begin{equation}
\label{djpfar}
\Pi_{ch}(y)={C}\left(\frac{1}{{y}^9} \mp
\frac{1}{{y}^3} + \frac{B}{{y}^4}\right)\cdot
\end{equation}

\begin{figure}
\includegraphics[width=0.8\linewidth]{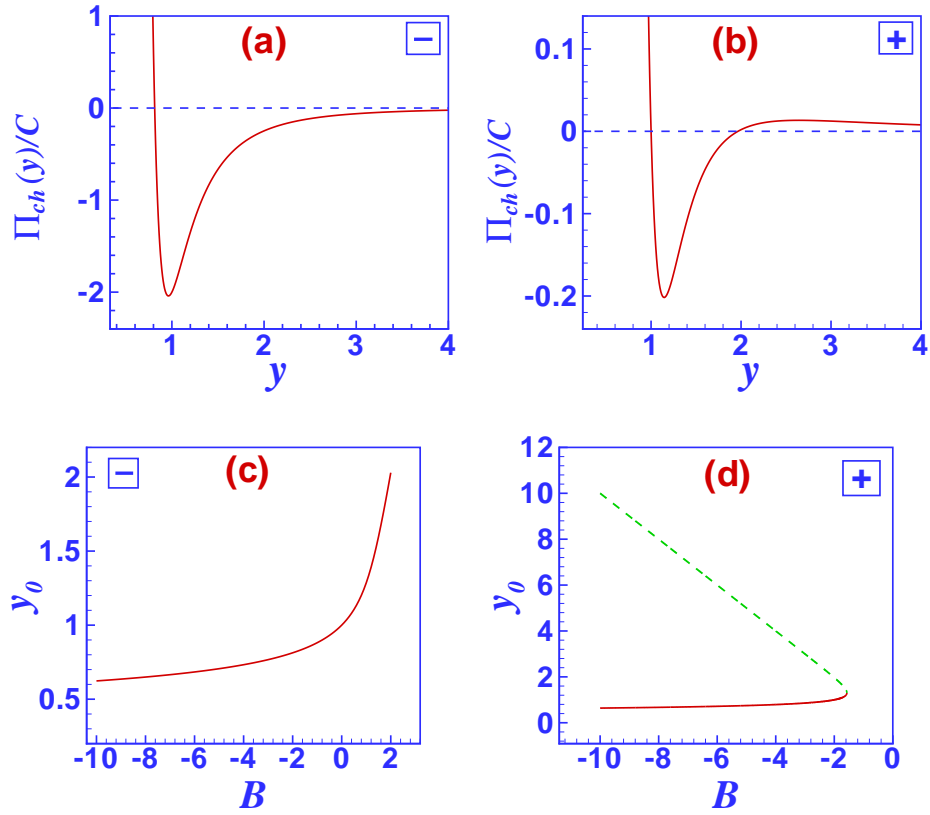}
\caption{Typical ($B=-2$) DJP (in units of $C$) of a flat
homogenous substrate for (a) the minus and (b) the plus case (see
Eq.~\eqref{djpfar})\/. The
corresponding zeros $y_0$ of the DJP for different values of $B$ are
given in (c) and (d) for the minus and the plus case,
respectively. In (c) and (d) full lines indicate stable wetting
films and dashed lines unstable films.}
\label{djpshapes}
\end{figure}

\begin{figure}
\includegraphics[width=0.8\linewidth]{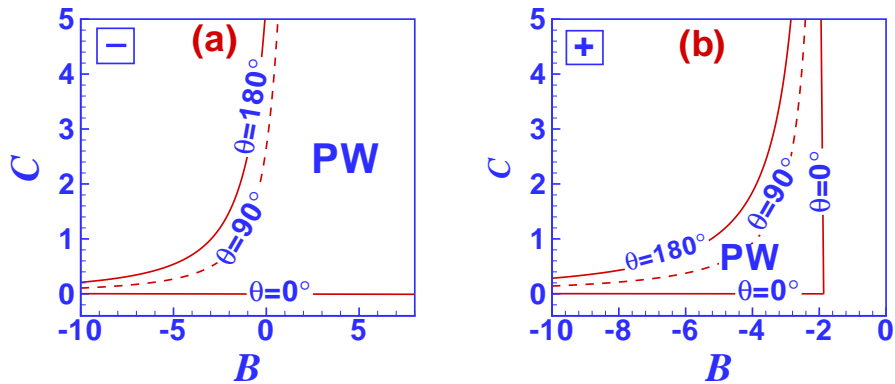}
\caption{ The value ranges of $B$ and $C$ for which the 
system exhibits a partial wetting (PW) situation, i.e.,
$0^\circ<\theq< 180^\circ$ for the minus (a) and the plus (b) case.}
\label{bc}
\end{figure}

Figure~\ref{djpshapes} shows the typical profile of $\Pi_{ch}(y)$
for the minus and the plus case and also the corresponding
equilibrium wetting layer thickness $y_0$ for which
$\Pi_{ch}(y_0)=0$\/. While the parameter
$C$ measures the strength of the DJP, by changing $B$ one can  
modify the shape of the DJP \cite{moosavi06b}\/.
%The parameters $C$ and $B$ allow one to change the strength and the shape of the DJP, respectively, for the minus and the plus case \cite{moosavi06b}. 
%equilibrium contact angle $\theta$ which can be calculated from \cite{dietrich88}
In Eq.~\eqref{djpfar} the admissible value ranges of $C$ and $B$ 
which provide partial wetting can be inferred from considering the 
equilibrium contact angle $\theta$ \cite{dietrich88}:
\begin{equation}
\label{eqtheta}
\cos\theta=1+\int_{y_0}^\infty
\Pi_{ch}(y)dy.
\end{equation}
%with $y_0$ as the equilibrium thickness of the wetting layer given 
% by the absolute minimum of the effective interface potential
% $\Phi$ (Eq.~\eqref{eflat}). 
The admissible value ranges of $B$ and $C$ for which $0^\circ<\theta<180^\circ$
(partial or incomplete wetting) are given in Fig.~\ref{bc} for both the minus and
the plus case. In the minus case, for each value of $B$ one can
find a value of $C$ 
such that the resulting substrate is partially wet.
Since the signs of the first two terms in Eq.~\eqref{djpfar} differ
the disjoining pressure has a zero for any $B$ and the depth of the
minimum of the corresponding effective interface potential can be
tuned by choosing an appropriate value for $C$\/. In the plus case,
however, $B$ has to be negative in order to obtain a sign change of
$\Pi$\/. 
The maximum admissible value of $B$ (i.e., $B<B_{max}$) can be obtained 
by simultaneously solving the following equations for $y_0$ and $B_{max}$:
%$B<B_{max}=-1.868$ for the plus case. $B_{max}$ corresponds to $\theq=0^\circ$ and 
%has been obtained by solving 
%\begin{subequations}
\begin{equation}
\Pi_{ch}(y_0)=\frac{1}{y_0^9}-\frac{1}{y_0^3}+\frac{B_{max}}{y_0^4}=0
\end{equation}
\begin{equation}
\Phi_{ch}(y_0)=\frac{1}{8y_0^8}-\frac{1}{2y_0^2}+\frac{B_{max}}{3y_0^3}=0,
\end{equation}
%\end{subequations}
from which one finds $B_{max}=-1.868$ (compare
Fig.~\ref{djpshapes}(d))\/.

In order to obtain dimensionless hydrodynamic equations
(see Eqs.~\eqref{eq:stokes}--\eqref{eq:surfacebc}) we choose
$A\,b/\mu$ as the velocity scale. With this, the dimensionless form of
the stess tensor is given by 
$\sigma_{ij}=-p\,\delta_{ij}+ C\,(\partial_j u_i+\partial_i u_j)$
and the surface tension coefficient drops out of
Eq.~\eqref{eq:surfacebc}\/.
The dimensionless time is given in units of $\mu/A$\/. In order to
study the dynamics of nanodroplets we solve the dimensionless
hydrodynamic equations with a standard biharmonic boundary integral
method described in more detail in \ref{numersec}\/.

%%%With the above modifications 
%%%%and the procedure of dimensionalization, 
%%%the DJP of the coated edge in the dimensionless form is given by
%%%\begin{eqnarray}
%%%\label{piend}
%%%\Pi^{\lhd}_{ce}(x,y)&=& C\bigg\{
%%%\frac{45\,I_e^{12}(x,y)}{\pi}
%%%\mp\frac{6\,I_e^{6}(x,y)}{\pi} \nonumber \\
%%%&&+\frac{2\,B\,[-I_c^{u}(x,y)-I_c^{u}(y,x)]}{\pi}\bigg\}.
%%%\end{eqnarray} 

%==========================   FIGURE  ================================== 
%=======================================================================
%==========================   FIGURE  ================================== 
%=======================================================================

\section{Results}
\label{results}
\subsection{Nanodroplets on homogeneous flat substrates}

%==========================   FIGURE  ================================== 
\begin{figure}
\begin{center}
\includegraphics[width=0.4\linewidth]{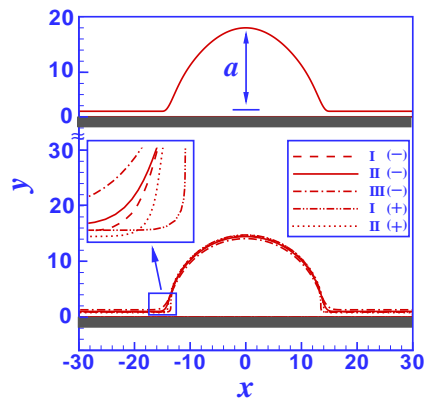}
\end{center}
\caption{Top panel: Droplet prepared in its initial configuration.
Bottom panel: The equilibrium configuration of a nanodroplet with 
$a=15$ on a flat homogeneous substrate for various values of $B$ 
and $C$ for the minus and the plus case. ($B$, $C$) for the minus 
case are I (-1, 7.7583) with $y_0=0.88$, II (0, 2.6667) with
$y_0=1$, and III (1, 1.2703) with $y_0=1.3$, and 
for the plus case I (-2.5, 4.2327) with $y_0=0.91$ and II (-4,
0.9265) with $y_0=0.79$\/. 
The values of $B$ and $C$ are chosen such that in all cases $\theq=90^\circ$.} 
\label{flat}
\end{figure}
%=======================================================================

In order to provide the information and terminology required
for the subsequent considerations we first recall
some basic results for the wetting of flat and homogeneous
substrates. For this purpose a nanodroplet with $a=15$ and an initial 
configuration given by Eq.~\eqref{inicond} was positioned on the 
substrate. Figure~\ref{flat} shows the equilibrium profile of the 
nanodroplet for various values of $C$ and $B$  
resulting in an equilibrium contact angle 
$\theq=90^\circ$ for both the minus and the plus case, i.e., the
values $(B,C)$ lie on the dashed curves in Figs.~\ref{bc}(a) and
\ref{bc}(b)\/. 
It is evident from the figure that the droplets have relaxed from
the initial condition. 
%and in order to equalize the pressure distribution inside the liquid phase the shape of the droplets has changed. 
The equilibrium profiles in all cases are roughly equal but the
nanodroplets differ near their contact lines (see the inset of
Fig.~\ref{flat}) and with respect to their heights\/. The term proportional to
$B$ in Eq.~\eqref{djpfar} is rather short-ranged and most important
in the direct vicinity of the substrate. The top parts of the droplets
are only influenced by the term $C/y^3$ such that the 
curvature at the peak changes with $C$, independently of $B$\/. This also
changes the droplet height. However, also the wetting film
thickness $y_0$ changes with $B$, such that the differences in droplet
height in Fig.~\ref{flat} are a combined result of both effects.

Due to the translational symmetry of the substrate and due to the
symmetry of the initial drop configuration the shape relaxation does not result in a lateral
displacement of the droplets, in contrast droplets placed  on
heterogeneous substrates \cite{ondarcuhu92,raphael88}\/.
%Because of the symmetry of the system there is no net lateral driving force on the droplets. 
%It can be shown that that by increasing (decreasing) the 
%size of the droplets the similarity will increase (decrease).

%The effect of the contact angle (keeping $B$ constant and 
%increasing value of $C$) on the equilibrium profile is shown in the 
%same figure. By increasing the equilibrium contact angle the contact 
%angle of nanodroplets defined, e.g., as their slope at the point of 
%inflection or via a spherical extrapolation of their top cap towards 
%the substrate increases. 

\subsection{Nanodroplets on the top side of steps}
Previous studies of droplets near edges (corresponding to steps of
infinite height) have shown that, in contrast to what is expected
from a simple macroscopic model taking into account only interface
energies, droplets are attracted towards the
edge in the minus case and repelled from the edge in the plus case
\cite{moosavi06b}\/. In the minus case, the droplets move towards
the edge with increasing velocity, but they stop rather abruptly
before the leading contact line reaches the edge. The distance from
the edge at
which the droplets stop increases with decreasing $B$, i.e., with
increasing strength of the coating layer. In the plus case, the
droplets move away from the step with a velocity which
decreases with the distance from the step. The strength of the
attraction or repulsion is expected to be lower for steps of finite height.

\subsubsection{Minus case}
%==========================   FIGURE  ================================== 
\begin{figure}
\includegraphics[width=0.8\linewidth]{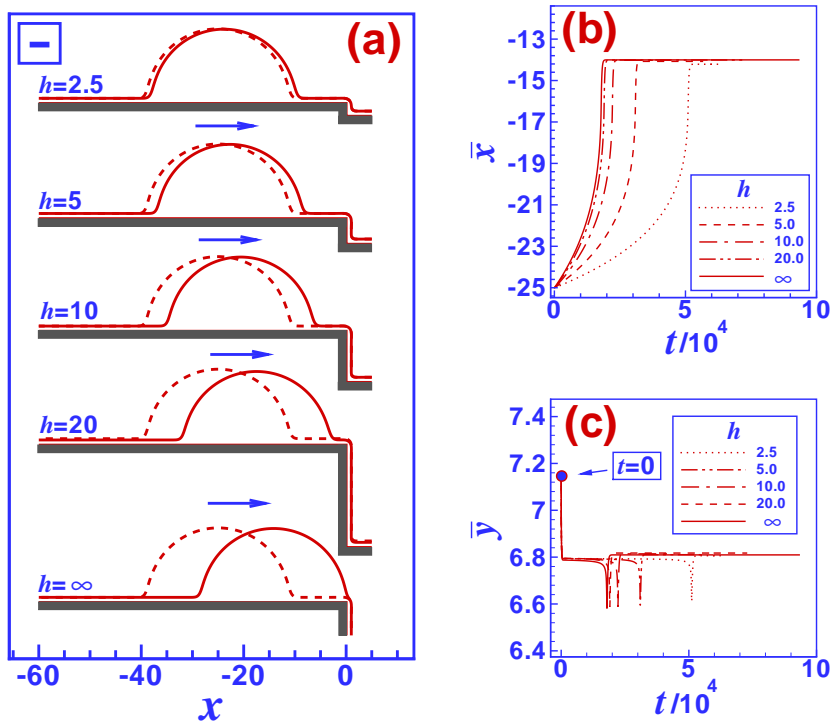}
\caption{
(a) The effect of the step height 
on the dynamics of nanodroplets positioned on the top side of a step for 
the minus case. The dashed and the solid lines correspond to 
nanodroplets at $t=200$ and $t=18700$, 
respectively. The droplets are initially positioned at a distance $\ell=10$ 
from the step. % (see Fig.~\ref{numfig}). 
The droplets have an initial radius  
$a=15$. $C=2.667$ and $B=0$ correspond to $\theq=90^\circ$\/. 
The vertical scale is equal to the lateral scale.
The corresponding lateral (b) and
vertical (c) position of the center 
of mass ($\bar{x}$, $\bar{y}$) of the droplet relative to the step
edge as a function of time.}
\label{effstepp}
\end{figure}
%=======================================================================
%==========================   FIGURE  ================================== 
\begin{figure}
\includegraphics[width=0.8\linewidth]{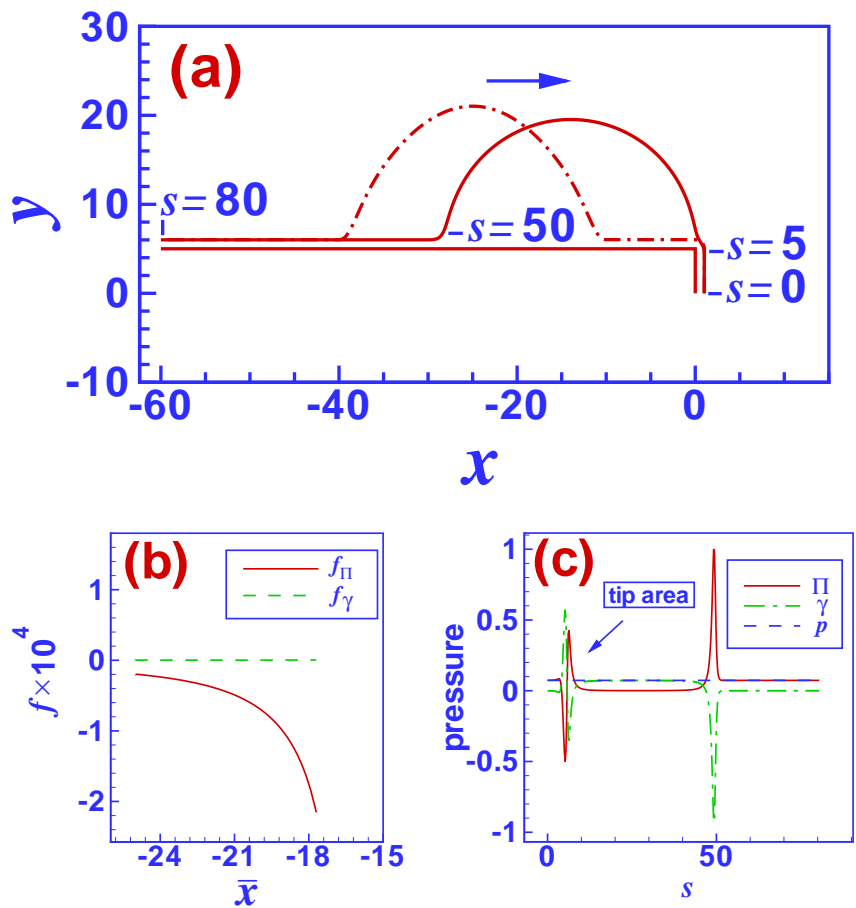}
\caption{(a) The initial and the final configurations of a droplet
with $a=15$\/. Initially it is positioned at $\ell=10$
($\bar{x}=-25$, dash-dotted line)\/. For the minus case it
moves towards an isolated edge where it stops with the leading
contact line pinned at the step edge (full line)\/. For this final
configuration the arclength $s$ of the interface is measured as indicated from a
certain position on the wetting layer on the vertical side of the
step.
(b) A comparison between
DJP induced (Eq.~\eqref{dpforce}, full line) and surface tension induced
(Eq.~\eqref{sigmaforce}, dashed line) lateral force densities
during the motion (with the leading three-phase contact line
still well separated from the edge)  expressed
in terms of the position $\bar{x}$ of the center of mass of the
droplet. 
(c) Laplace pressure $\gamma\,\kappa$ (dash-dotted line) and DJP
$\Pi$ (full line) on the surface
of the droplet in the final equilibrium configuration as a function of the
arclength $s$\/. In the absence of other external forces (e.g., $g$
in Eq.~\eqref{eq:surfacebc}) at each point on the droplet surface 
these add up to the constant pressure $p$ inside the droplet.
$B=0$ and $C=2.667$ correspond to $\theq=90^\circ$\/. Force
densities (b) and
pressures (c) are measured in units of $\gamma/b^2$ and $\gamma/b$, respectively.}
\label{comppisigma}
\end{figure}
%=======================================================================

%==========================   FIGURE  ================================== 
\begin{figure}
\includegraphics[width=0.8\linewidth]{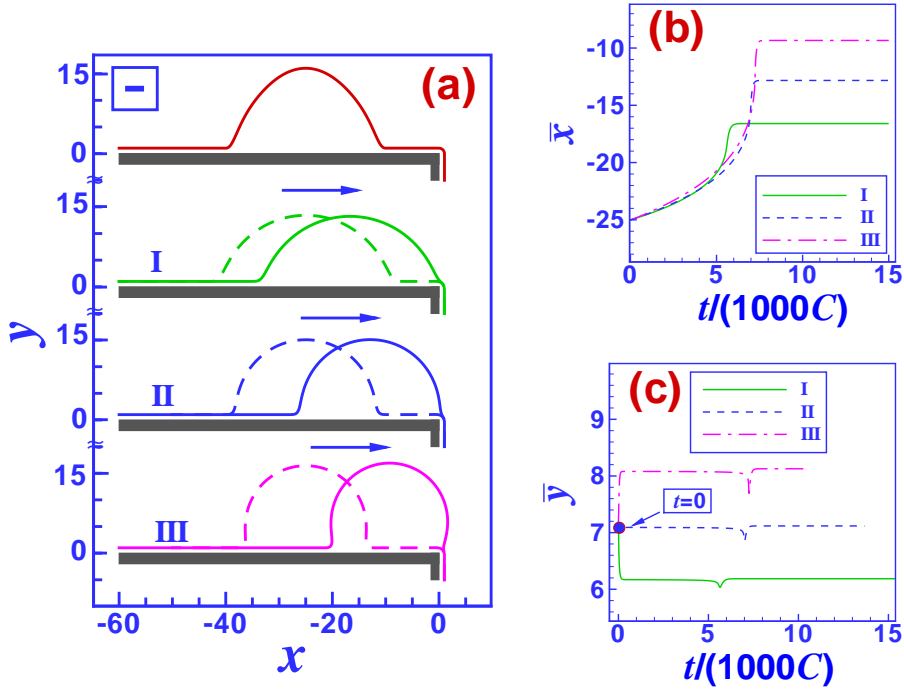}
\caption{\label{edgeeffc} The effect of the contact angle on the
dynamics of droplets on the top side of the step for a droplet with
initial height $a=15$ and for the minus case.  (a) The  initial profile
$(\ell=10)$ is shown in the top panel. The lower graphs show the
configurations of the droplets after the initial relaxation
($t/C\approx 60$, dashed lines) and in the final stages ($t/C=$6200,
7400, and 7800 for I, II, and III, respectively, solid lines) for
$\theq=75.5^\circ$ (I), $97.2^\circ$ (II), and $120^\circ$ (III)
from top to bottom. The corresponding substrate parameters are I
$(C=2, B=0)$, II $(C=3, B=0)$, and III $(C=4, B=0)$, respectively.
In (b) and (c) as function of time the corresponding lateral and vertical positions $\bar{x}$ and
$\bar{y}$, respectively, of the center of mass
of the droplets are shown relative to the step edge. The dips in
(c) occur when the leading three-phase contact line reaches the
edge; then the droplet stops (compare to (b))\/. }
\end{figure}
%=======================================================================
%==========================   FIGURE  ================================== 
\begin{figure}
\includegraphics[width=0.8\linewidth]{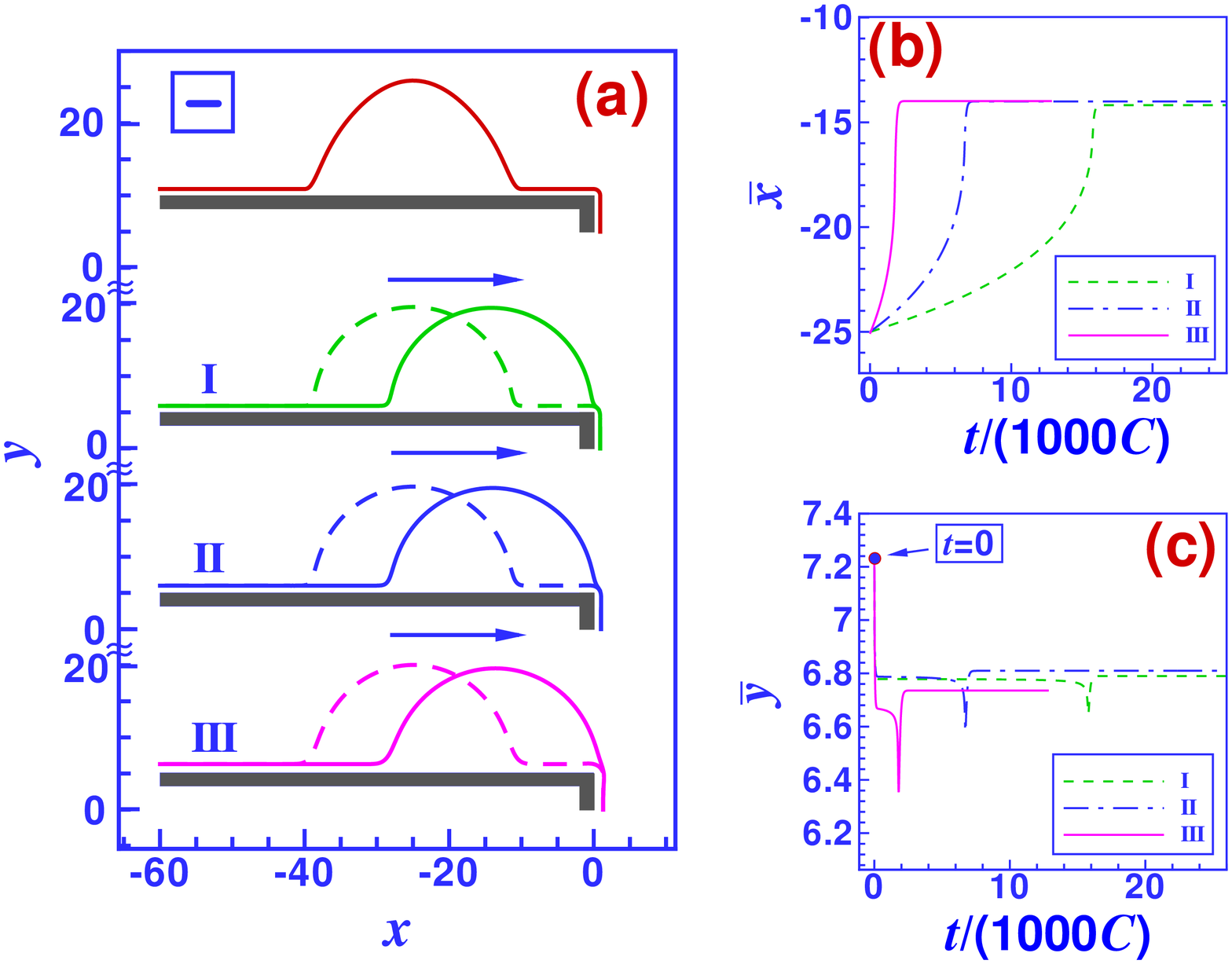}
\caption{The effect of changing $B$ (while varying $C$ 
such that $\theq=90^\circ$ in all cases, see Fig.~\ref{bc}(a)) on
the dynamics of the droplets on the top side
of an edge for an initial droplet with $a=15$ and for the 
minus case. The values of $B$ and $C$ are I ($B=-1$, $C=1.2703$), 
II ($B=0$, $C=2.6667$), and III ($B=1$, $C=7.7583$)\/. 
(a) The top panel depicts the initial profile
($\ell=10$)\/. 
The dashed lines show the configuration of the droplet after the initial 
relaxation $t/C\approx 60$ and the solid lines correspond to the 
final configuration of the droplets at $t/C=16400$, 7300, and 2500 for I, II,
and III, respectively. In (b) and (c) as a function of time the corresponding lateral and
vertical positions $\bar{x}$ and $\bar{y}$, respectively, of the
center of mass of the droplets are shown relative to the
step edge.
%Times is scaled by $C$\/.% \cite{moosavi08a}.
}
\label{edgediffb}
\end{figure}
%=======================================================================
%==========================   FIGURE  ================================== 
%%%%\begin{figure}[b]
%%%%\includegraphics[width=0.4\linewidth]{10x}
%%%%\includegraphics[width=0.4\linewidth]{10y}
%%%%\caption{The position of the center of mass of the droplets 
%%%%($\bar{x}$, $\bar{y}$) for nanodroplet with $a=10$ residing near an edge for 
%%%%$\theq=75.5^\circ$ (I), $97.2^\circ$ (II), and $120^\circ$ (III). 
%%%%The contact angles correspond to  I $(C=2, B=0)$, II $(C=3, B=0)$, 
%%%%and III $(C=4, B=0)$, respectively. In order to compare the profile the times are scaled by $C$ values.
%%%%%\cite{moosavi08a}.
%%%%}
%%%%\label{edgeeffcc}
%%%%\end{figure}
%=======================================================================

In order to test the influence of the step height on the dynamics
of nanodroplets 
identical droplets of half base width $a=15$ were placed at a
distance $\ell=10$ from steps
%, as depicted in Fig.~\ref{numfig} (only consider the droplets on
%top of the step), with 
of height $h=$2.5, 5, 10, 15, 20, and $\infty$\/. The results of the 
numerical solution of the mesoscopic hydrodynamic equations 
for the minus case are shown in Fig.~\ref{effstepp}(a) 
for $C=2.667$ and $B=0$ which corresponds to $\theq=90^\circ$\/. 
In order 
to have a better view on the dynamics we monitor the time evolution 
of the position of the center of mass of the droplets ($\bar{x}$, $\bar{y}$)
relative to the step edge
in Figs.~\ref{effstepp}(b) and \ref{effstepp}(c), where $\bar{x}$ and $\bar{y}$ 
are given by
\begin{equation}
\label{ave}
%\bar{x}=\frac{\int_{\Gamma_d}\,x\,ds}{\int_{\Gamma_d}\,ds},\quad 
%\bar{y}=\frac{\int_{\Gamma_d}\,y\,ds}{\int_{\Gamma_d}\,ds}
\bar{x}=\frac{\int_{\Omega_d}\,x\,dV}{\int_{\Omega_d}\,dV},\quad 
\bar{y}=\frac{\int_{\Omega_d}\,y\,dV}{\int_{\Omega_d}\,dV}-h,
\end{equation}
with $\Omega_d$ denoting the droplet volume. 
Since the droplets are smoothly connected to
the wetting film, which on large substrates would influence the
center of mass of the fluid, in calculating $\bar{x}$ 
and $\bar{y}$ we only consider the fluid above $y=c_0\,y_0+h$ with
$c_0>1$, i.e., only the fluid volume slightly above the wetting
film. We selected $c_0=1.2$; but since we focus on substrates with
equilibrium contact angles of about $90^\circ$ the results are only
weakly affected by the precise choice of the value of $c_0$\/.

In all cases the dynamics of the droplets proceeds in three stages.
The first stage is a fast initial shape relaxation, similar to the
behavior on homogeneous substrates, which is accompanied by a
lowering of the droplet center of mass $\bar{y}$ without any
considerable lateral motion. This is followed by a relatively
slow lateral motion towards the edge, during which the changes in
the droplet shape are almost unnoticeable. Although the 
droplet shape is slightly asymmetric the lateral surface tension
induced force density $f_{\gamma}$ defined in Eq.~\eqref{sigmaforce}
is much smaller than the force density $f_{\Pi}$  induced by
the DJP (see Eq.~\eqref{dpforce}) as shown in
Figs.~\ref{comppisigma}(b)\/.
%As can be seen for all the cases considered the droplets have 
%experienced two different reactions. First, similar to the case 
%of the droplets on flat and homogeneous substrates a fast initial 
%relaxation, namely, change of the position of the center of mass of 
%the droplets in the vertical direction ($\bar{y}$) without any 
%considerable lateral motion. Secondly a lateral motion towards the 
%edge accompanied with a slight relaxation, because of dissimilarity 
%of the disjoining pressure on the droplet sides which provides a 
%driving force. Although the droplet shape is not symmetric in this 
%stage, the lateral force $f_{\gamma}$ because of the surface tension is much 
%smaller than that induced by the DJP as is shown in Fig.~\ref{comppisigma}(a) and \ref{comppisigma}b . 
Figure~\ref{effstepp}(b) clearly shows that the lateral motion of the
droplet slows down rapidly as soon as its leading three phase
contact line
reaches the edge. During this third and final stage a part of the droplet
volume leaks into the wetting film on the vertical part of the step
and as a result the droplet
experiences a sudden drop in its height $\bar{y}$ (see
Fig.~\ref{effstepp}(c))\/. The trailing three-phase contact line of
the droplet still
continues its motion towards the step and as a consequence the mean
height of the droplet increases again and becomes even larger than
during the migration stage. While the droplet contracts, its 
asymmetry gradually increases such that the surface tension induced
force density
$f_{\gamma}$ grows and finally, as the equilibrium configuration is
reached, cancels $f_{\Pi}$\/. (This latter stage of cancelation 
is not visualized in
Fig.~\ref{comppisigma}(b) due to numerical problems in evaluating
the force densities on droplets once they have reached the step
edge.)
In equilibrium, at each point on its
surface the Laplace pressure
and the disjoining pressure ad up to the constant value of the 
hydrostatic pressure in
the droplet (see Fig.~\ref{comppisigma}(c))\/. 

Increasing the step height from $h=2.5$
to 5 and 10 results in a significant increase in the droplet speed during
the migration phase. The asymptotic speed for
isolated edges (corresponding to $h=\infty$), i.e., the maximum
speed, is almost reached for $h=20$\/. 
%The other feature is that the rate of decreasing the speed with
%distance from the step depends on the step size. Contour lines of
%the DJP are flattened faster in the case of small step sizes and,
%as results, step size controls the rate of reduction of the
%droplets speed. 
This height value is large compared to the thickness of the
wetting layer but comparable with the droplet size; here the base
diameter is $2\,a=20$\/.
However, in order to be able to conclude that the step height above which the
droplet perceives the step as an isolated edge is comparable with the
droplet size further calculations for droplets of different size
are needed.

Changing the equilibrium contact angle $\theq$ by increasing
$C$ while keeping $B=0$ does not qualitatively change the behavior of the
droplets, as shown in Fig.~\ref{edgeeffc}(a) for droplets with $a=15$
near an isolated edge (corresponding to $h=\infty$), apart from the
increase of $\bar{y}$ during the initial relaxation process for large
$\theq$\/. The reason for this increase in droplet height 
is, that the initial shape of
the droplet is not adapted to the substrate parameters.
Changing $C$ does not change the functional form of the DJP, only
its
strength. Consequently, droplets move faster for larger $C$
(resulting in larger $\theq$) and their final shape is less
symmetric. With the leading contact line pinned right at the step
edge, large $\theq>90^\circ$ also result in an overhang over the
step edge. Since for fixed wetting film thickness $b$ on the
uncoated flat substrate the time scale used to obtain dimensionless
hydrodynamic equations depends on the substrate parameters in the
same manner as the dimensionless parameter $C$, we rescale time by
$C$ in Figs.~\ref{edgeeffc}(b) and
\ref{edgeeffc}(c), as well as in all subsequent figures which
compare
$\bar{x}(t)$ and $\bar{y}(t)$ for different values of $C$\/.
This corresponds to changing the substrate material but
keeping surface tension, viscosity, and wetting film thickness
constant \cite{moosavi08a}\/.

%The influence of the contact angle $\theq$ (keeping $B=0$ and
%increasing $C$) on the dynamics and the final configuration of the
%droplets is shown in Fig~\ref{edgeeffc}a for a droplet with $a=15$
%near an isolated edge (corresponding to $h=\infty$)\/.
%In orther to properly comparing the dynamics of the cases the
%times of each case are scaled by $C$ value of that case assuming
%that $b$ is equal for all the cases considered \cite{moosavi08a}.
%A larger value of the contact angle although does not change the
%form of the distribution of the DJP over the step but increases
%the strength of DJP and imposes a stronger lateral induced force
%on the droplets. As a results the droplets move faster and have
%been displaced more towards the step (see Figs.~\ref{edgeeffc}b
%and \ref{edgeeffc}c).
%%%gradient of the DJP althoth the distribution of the DJP ( remains the same. 
%%%As a result a largere lateral induced force is provided and the droplets have 
%%%%%been displaced more towards the step. 
%%In order to provide the required 
%%surface tension forces to stop the droplet, the leading edge of
%the droplets have been also 
%deformed more over the tip area for larger values of $C$. 
%%%The corresponding time dependent positions of the center of mass of the 
%%%droplets ($\bar{x}$, $\bar{y}$) are shown depicted in \ref{edgeeffcthe same figurer 
%%%[see Fig.~\ref{edgeeffc}(b) and (c)]. 

Figure \ref{edgediffb}(a) shows the effect of changing the value of
$B$ (while keeping the contact angle constant) on the dynamics and
on the final configuration of droplets which start with $a=15$ and
for the minus case on the top side of an
isolated edge ($h=\infty$)\/. 
%[I ($B=-1$, $C=1.2703$), II ($B=0$, $C=2.6667$), and III ($B=1$, $C=7.7583$)]. 
For each value of $B$ we choose 
$C$ such that $\theq=90^\circ$,
i.e., corresponding to the dashed curve in Fig.~\ref{bc}(a)\/. 
%In orther to properly comparing the dynamics of the cases the times of each case are scaled by $C$ value of that case assuming that $b$ is equal for all the 
%cases considered \cite{moosavi08a}.
For all values of $B$ the droplets move towards the edge. 
Changing the values of $B$ and $C$ does not qualitatively change the behavior 
of the system. However, a closer examination of $\bar{x}$ and 
$\bar{y}$ (see Figs.~\ref{edgediffb}(b) and
\ref{edgediffb}(c), respectively) reveals quantitative differences in
the dynamics and in the final configuration of the droplets despite
the fact that the contact angle is the same for all these cases.
For larger (positive) values of $B$ (and thus $C$, see
Fig.~\eqref{bc}(a)) 
droplets move faster in lateral direction
although the contact angle equals $\theq=90^\circ$ for all of them. In addition, the
final position of the droplets is closer to the step edge for
larger values of $B$, eventually leading to a slight overhang. The
small differences in $\bar{y}$ for different values of $B$ is related to
the fact that the shape of nanodroplets is not only determined by
$\theq$, as shown in Fig.~\ref{flat}, and that the wetting film
thickness depends on $B$\/. 

\subsubsection{Plus case}
%==========================   FIGURE  ================================== 
\begin{figure}
\includegraphics[width=0.8\linewidth]{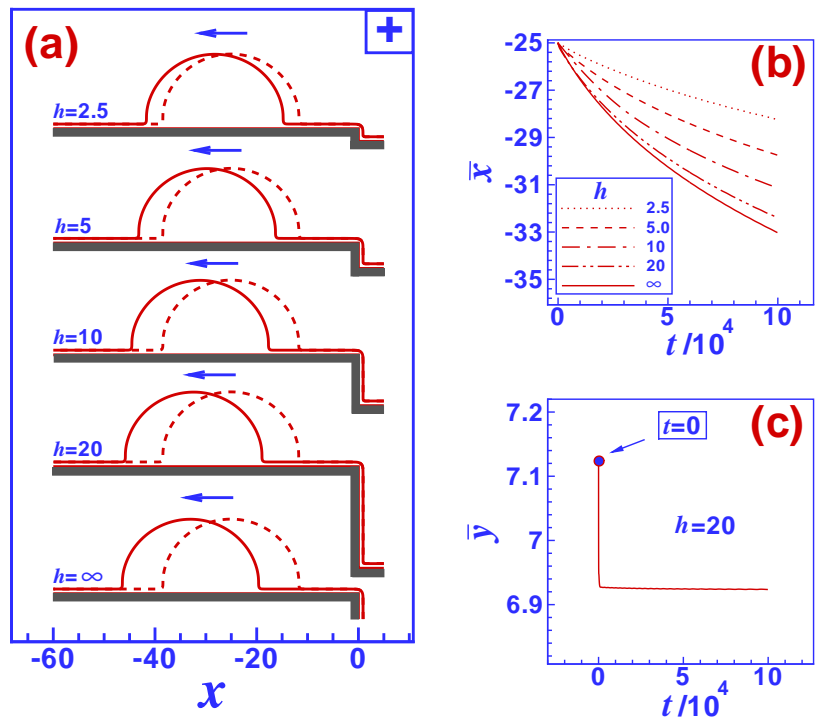}
\caption{ %Nanodroplets residing up and near a step are repeleld from the step in the plus case. 
(a) The effect of the step height 
on the dynamics of nanodroplets positioned on the top side of a step for 
the plus case. The dashed and the solid lines correspond to 
times $t=200$ and $t=10^5$, respectively. The droplets of size
$a=15$ are initially positioned at a distance $\ell=10$ from the
step.
% (see Fig.~\ref{numfig})\/. 
$C=4.2327$ and $B=-2.5$ result in an equilibrium contact angle 
$\theq=90^\circ$\/. Time evolution of the (b) horizontal position
$\bar{x}$  and (c) vertical
position $\bar{y}$ of the center of mass of the droplets relative to the
step edge.
Since $\bar{y}$ depends only weakly on $h$ only the case $h=20$ is
shown.
}
\label{effstepplus}
\end{figure}
%==========================   FIGURE  ================================== 
\begin{figure}
\includegraphics[width=0.8\linewidth]{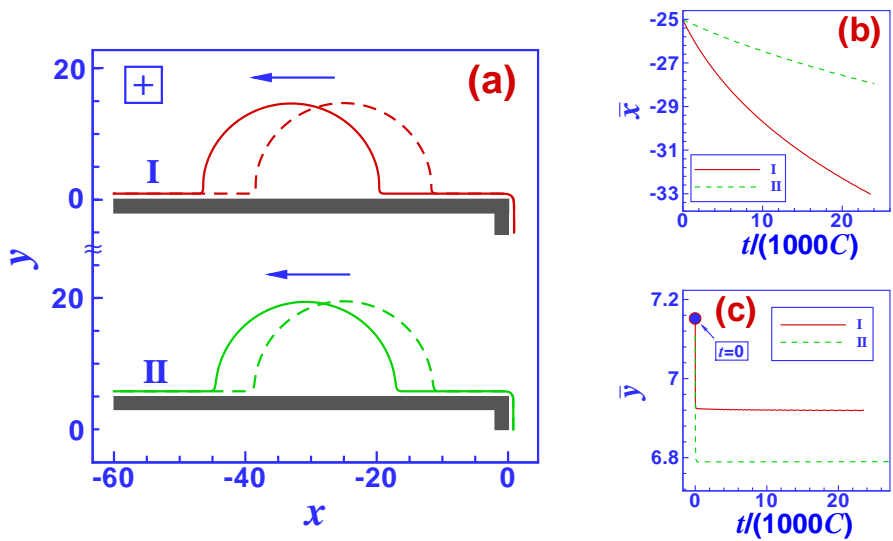}
\caption{The effect of $B$ on the dynamics of the droplets 
near an edge for the plus case and for an initial droplet shape with $a=15$\/.
The values of $B$ and $C$, i.e., I ($B=-2.5$, $C=4.2327$) 
and II ($B=-4$, $C=0.9265$) are selected such that the equilibrium 
contact angle is $\theq=90^\circ$ in both cases. (a) 
The initial distance from the step edge is $\ell=10$\/. 
The dashed lines show the droplets 
after the initial relaxation at $t/C\approx60 $  and the solid lines correspond 
to the droplets in the migration phase at $t/C=24000$\/. 
The horizontal position $\bar{x}$  and the vertical position $\bar{y}$ of the center 
of mass of the droplet are shown as a function of time in (b)
and (c), respectively. For less negative values of $B$ the velocity
$d\bar{x}/dt$ is larger (I)\/.
%In order to compare the profile the times are scaled by $C$ values.
%\cite{moosavi08a}.
\label{edgeplus}}
\end{figure}
%==========================   FIGURE  ================================== 

Even if they exhibit the same equal equilibrium contact angles $\theq$ the behavior of
droplets in the plus case differs substantially from that in the minus
case: the direction of motion is reversed. However, apart from this
sign change, the influences of the step height, the equilibrium contact
angle, and $B$ are similar.

The dependence of the droplet dynamics on the step height for the
plus case is shown in Fig.~\ref{effstepplus}(a)\/.
The initial size of the droplets is $a=15$ and the contact angle
$\theq=90^\circ$ (with $C=4.2327$, $B=-2.5$)\/. The corresponding 
lateral and vertical positions of the center of mass of the
droplets relative to the step edge are shown in
Figs.~\ref{effstepplus}(b) and \ref{effstepplus}(c), respectively.
As in the minus case the migration phase is preceded by a fast
initial relaxation process (during which $\bar{y}$ drops
slightly)\/. However, the droplets are repelled from the step.
The lateral speed of the motion continuously 
decreases as the distance of the droplets from the step edge increases.
For higher steps the droplets are faster. But as in the minus
case, the maximum speed (reached for $h=\infty$, i.e., in the case
of an isolated edge) is almost reached for $h= 20$ (see
Fig.~\ref{effstepplus}(b))\/.

The results 
for different values of $B$, while keeping the contact angle
$\theq=90^\circ$ fixed,
%[I ($B=-2.5$, $C=4.2327$) and II ($B=-4$, $C=0.9265$)] 
are depicted in Fig.~\ref{edgeplus}(a)\/. 
The corresponding lateral $\bar{x}$ and vertical $\bar{y}$
positions of the center of mass of the droplet relative to the
step edge are given in Figs.~\ref{edgeplus}(b) and
\ref{edgeplus}(c), respectively. For all
the cases considered the droplets move away from the step. However,
as in the minus case, the droplet speed increases with $B$ (i.e.,
for less negative values of $B$), even
though the contact angle is not changed. The reason for this is,
that larger (i.e., less negative) values of $B$ require larger
values of $C$ in order to maintain the same $\theq$\/. As in the
minus case the droplet height, i.e., $\bar{y}$, depends on $B$ as well. 
%the form of the distribution of the DJP
%over the step provide conditions in which the dynamics is enhanced
%and the speed of the droplets is larger for the cases with more
%positive value of $B$. 

%This again , since more positive values of $B$ corresponding to larger values of $C$ provide a 
%larger force on the nanodroplets, the dynamics is enhanced and the speed of the 
%droplets is larger even though the parameters ($C$ and $B$) are selected such that the contact angle is equal for all the cases.

\subsection{Nanodroplets at the step base}

In Ref.~\cite{moosavi06b} we have demonstrated that droplets near
corners,
i.e., at the base of a step of infinite height, are attracted to
the corner in the plus case and repelled from the corner in the minus case (while in
a macroscopic model taking into account only interface energies the
free energy of the droplets is independent of their distance from
the corner)\/.  In other
words, the direction of motion is reversed as compared to the case
of the edge. However, as we shall show in the following, at a step
composed of an edge and a corner at its base, the direction of
motion of nanodroplets is the same on both sides of the step.

%Previous studies of droplets near edges (corresponding to steps of
\subsubsection{Minus case}
%===========%==========================   FIGURE  ==================================
\begin{figure}[b]
\includegraphics[width=0.8\linewidth]{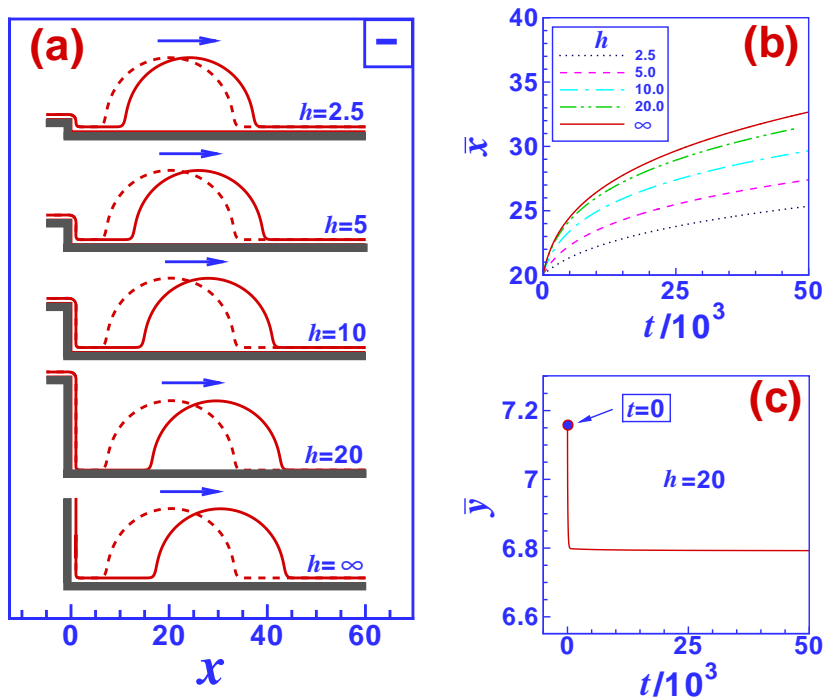}
\caption{
(a) Nanodroplets with initial size $a=15$ positioned at the base of
topographic steps of different height for the minus case. The
droplets start at a distance $\ell=10$ from the step. $C=2.667$ and
$B=0$ correspond to $\theq=90^\circ$\/. The dashed and the solid
lines correspond to the configurations just after the initial relaxation
at $t=170$ and to a later time $t=30000$, respectively.  As
function of time the
horizontal position $\bar{x}$ and the vertical position $\bar{y}$ of the center
of mass relative to the step base are shown in (b) and (c),  respectively.
Since $\bar{y}$ depends only weakly on $h$, in (c) only the
trajectory for $h=20$ is shown.}
\label{wedgediffhminus}
\end{figure} 
%==========================   FIGURE  ==================================
\begin{figure}
\includegraphics[width=0.8\linewidth]{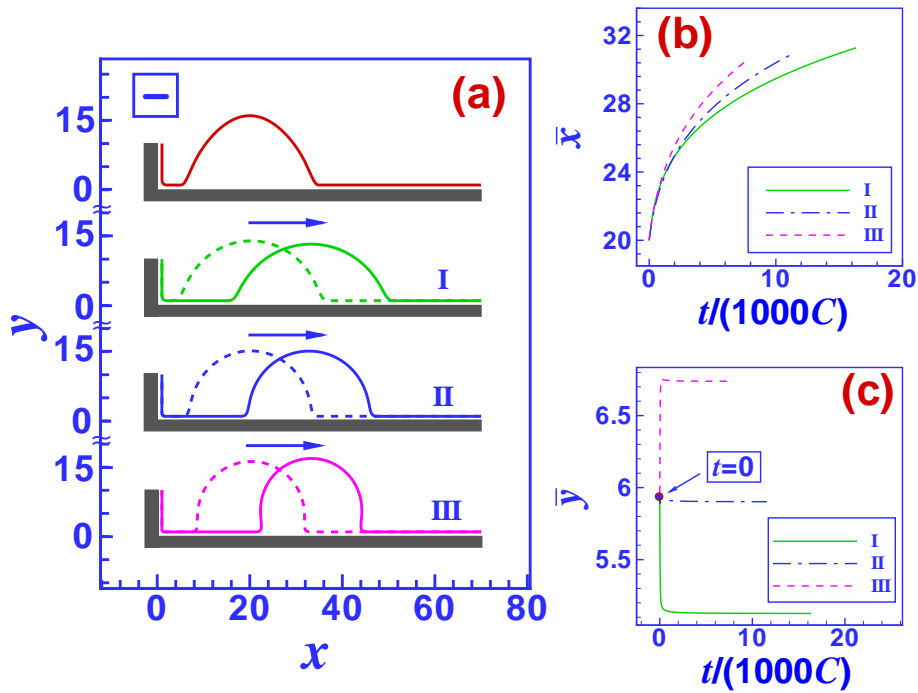}
\caption{ (a) Droplets of initial size $a=15$ near a corner of
substrates with different contact angles $\theq$ in the minus case. 
The top panel shows the initial droplet profile with $\ell=5$\/. 
The other panels show the droplets after the initial 
relaxation ($t/C=44.5$ (I), $20.3$ (II), and $34.25$ (III), dashed lines) and
during the migration stage ($t/C=26100$ (I), $17200$ (II), and
$13600$ (III), solid lines) 
for $\theq=75.5^\circ$ (I, $C=2$, $B=0$), $97.2^\circ$ (II, $C=3$,
$B=0$), and $120^\circ$ 
(III, $C=4$, $B=0$), respectively. 
The time evolution of the center of mass $(\bar{x}, \bar{y})$ relative to
the corner is shown in (b) and (c) for the lateral and vertical
direction, respectively.
%In order to compare the profile the times are scaled by $C$ values.
%\cite{moosavi08a}.}
\label{wedgediffcminus}}
\end{figure}
%==========================   FIGURE  ==================================
\begin{figure}[b]
\includegraphics[width=0.8\linewidth]{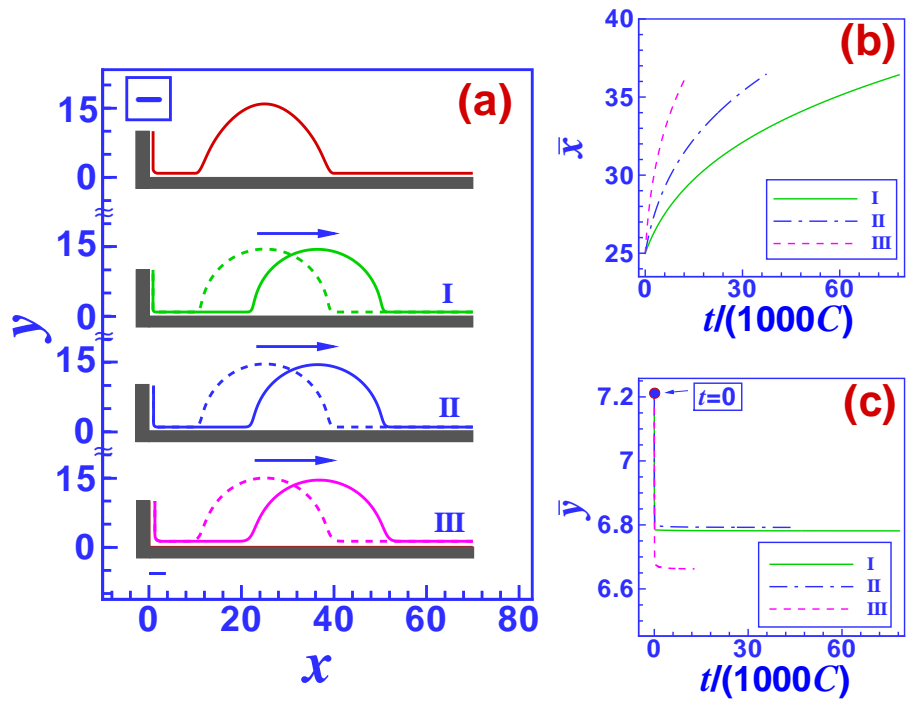}
\caption{(a) Nanodroplets on substrates with different $B$ but the same
contact angle $\theq=90^\circ$ near a corner in the minus case:
(I: $B=-1$, $C=1.2703$), (II: $B=0$, $C=2.6667$), and (III:
$B=1$, $C=7.7583$), top to bottom\/. The top panel depicts the
initial droplet shape $(\ell=10)$\/. The corresponding graphs show the
droplets after the  initial relaxation
(dashed lines, $t/C\approx 60$) and in the migration process at
$\bar{x}=36.5$  (solid lines,  $t/C=78500$, 37500, 
and 12700 for I, II, and III, respectively)\/.
The time evolution of the center of mass $(\bar{x},\bar{y})$
relative to the corner is shown in (b) and (c) for the lateral and
vertical direction, respectively.
%In order to compare the profile the times are scaled by $C$ values.
%\cite{moosavi08a}.}
\label{wedgediffb}}
\end{figure} 
%==========================   FIGURE  ==================================
\begin{figure}
\begin{center}
\includegraphics[width=0.4\linewidth]{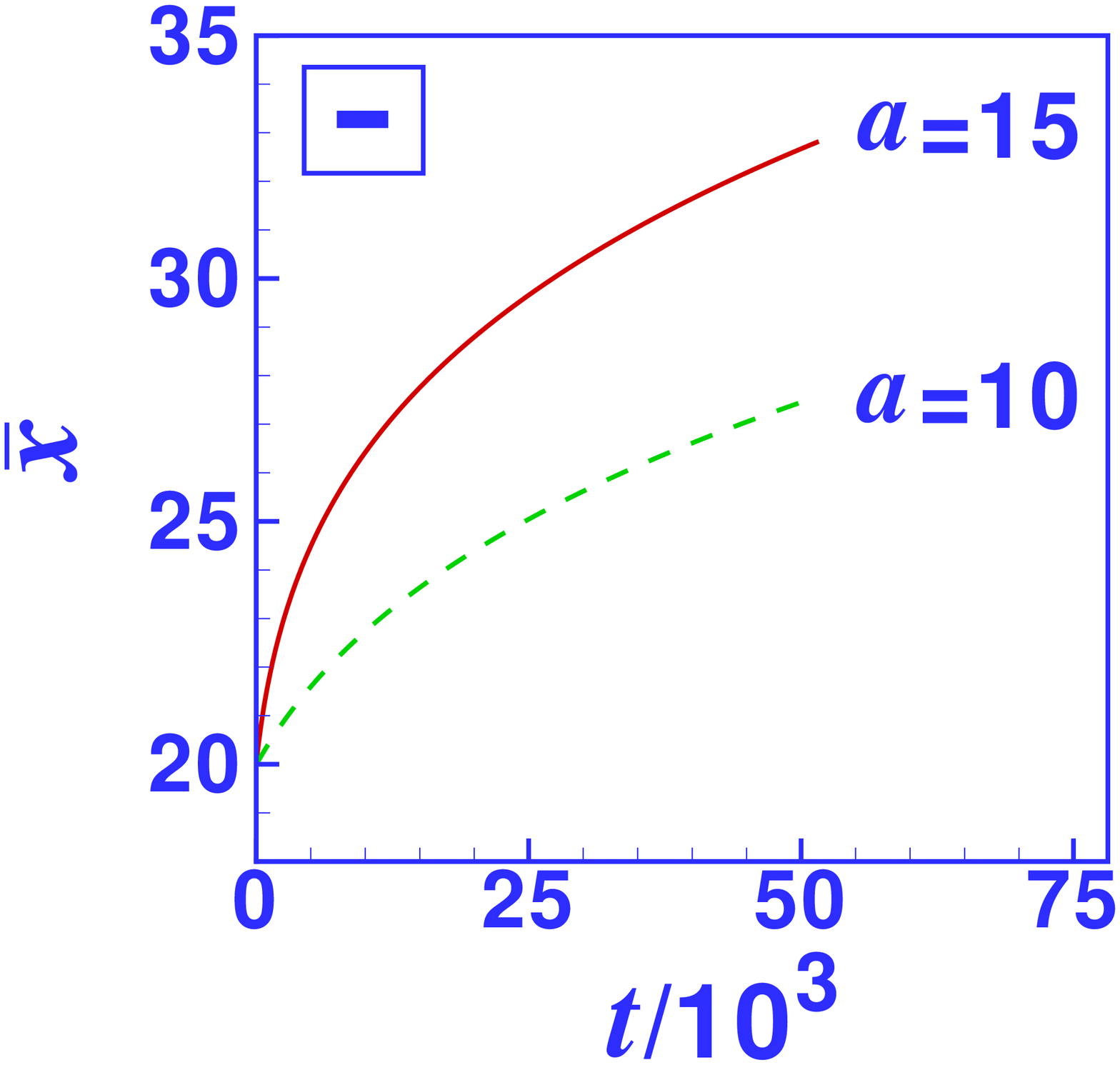}
\end{center}
\caption{The lateral position $\bar{x}$ of the center of mass as a
function of time for droplets of size $a=15$
(solid line) and $a=10$ (dashed line) 
near the corner of a wedge in the minus case with $B=0$,
$C=2.6667$, and $\theq=90^\circ$\/.  }
\label{wedgedropsize}
\end{figure}
%=======================================================================

As in the case of droplets on the top side of steps, the step height
influences the droplet velocity but not the direction of motion and
the transition from a planar substrate ($h=0$) to an isolated wedge
($h=\infty$) is continuous. This is demonstrated in
Fig.~\ref{wedgediffhminus} for droplets of
size $a=15$ starting at a distance $\ell=10$ from the corner.
The initial distance $\ell$ is chosen such that after the initial relaxation which
precedes the migration phase the contact line
facing the corner is well separated from the wetting layer on the
vertical part of the step. For the minus case
Fig.~\ref{wedgediffhminus}(a) presents
the results of our Stokes dynamics calculations for droplets at the step base
for different step heights.
The droplets are repelled from the step and move away 
with a speed which decreases continuously with the distance from the step. 
The differences in droplet speed are significant between step heights
$h=2.5$, 5, and 10 (see Fig.~\ref{wedgediffhminus}(b))\/. Increasing
the step height further influences the dynamics of the droplets
only at large distances from the step.

Changing the equilibrium contact angle $\theq$ does not change the
droplet dynamics qualitatively. This is demonstrated in
Fig.~\ref{wedgediffcminus}(a) for droplets on substrates with
different values of $C$ (while keeping $B=0$)\/. The top panel
shows the initial shape used in all cases considered here for the
numerical solution of the Stokes dynamics. 
The corresponding lateral position $\bar{x}$ and vertical position $\bar{y}$
of the center of mass of the droplets relative to the
corner are depicted in
Figs.~\ref{wedgediffcminus}(b) and \ref{wedgediffcminus}(c),
respectively. Increasing $\theq$ (by increasing $C$) results in
faster droplet motion. Since the initial droplet shape is not adapted to the
modified contact angle $\theq\ne 90^\circ$, $\bar{y}$ changes rapidly during the
initial relaxation process for $\theq\ne 90^\circ$\/.

The dynamics of droplets on substrates with the same contact angle
$\theq=90^\circ$ but
different values of $B$ (with $C$ adapted accordingly) is shown in 
Fig.~\ref{wedgediffb}(a)\/. The top panel shows the initial
configuration.
%namely, ($C=1.27$, $B=-1$), ($C=2.667$, $B=0$), and 
%($C=7.76$, $B=1$), 
%All the case provide $\theq=90^\circ$. 
For all cases considered the droplets move away from the step.
Comparing $\bar{x}(t)$ for different values of $B$ (see
Fig.~\ref{wedgediffb}(b)) shows, that the droplet velocity increases
with $B$ (which, for fixed $\theq=90^\circ$,
implies increasing $C$)\/. After the initial
relaxation process the vertical coordinate $\bar{y}$ of the center of mass
does not vary in time (see
Fig.~\ref{wedgediffb}(c))\/.

The droplet dynamics depends on the droplet size. This is demonstrated
in Fig.~\ref{wedgedropsize} for droplets of initial sizes $a=10$ and $a=15$
starting at $\bar{x}=20$ near an isolated wedge for the minus case. 
%Although both the droplets are 
%initially positioned at $\bar{x}=20$ but 
The larger droplet moves faster because its two three-phase contact
lines have a larger lateral distance from each other such that they experience a
larger difference in the local disjoining pressure.
%Moreover, close to the corner of the wedge, the trailing side of the droplet is
%rather close to the vertical part of the wedge and experiences an
%additional repulsive force which increases proportional with the droplet height.
%As shown in Fig.~\ref{dpstep}(a), the disjoining pressure above the
%wetting film is more negative on the left side than
%on the right hand of the droplet.

%because of the difference in the height of the 
%droplets the upper part of the larger droplet can provide an additional 
%driving force. Considering that a flat substrate with a lateral 
%variation of DJP can be considered as a chemically structured substrate with 
%a continuous wettability gradients the results are in agreement 
%with findings of Ref.~\cite{subramanian05} on droplets on 
%such substrates that the speed of the droplets is proportional to the size of the droplets 
%

\subsubsection{Plus case}
%==========================   FIGURE  ==================================
\begin{figure}
\includegraphics[width=0.8\linewidth]{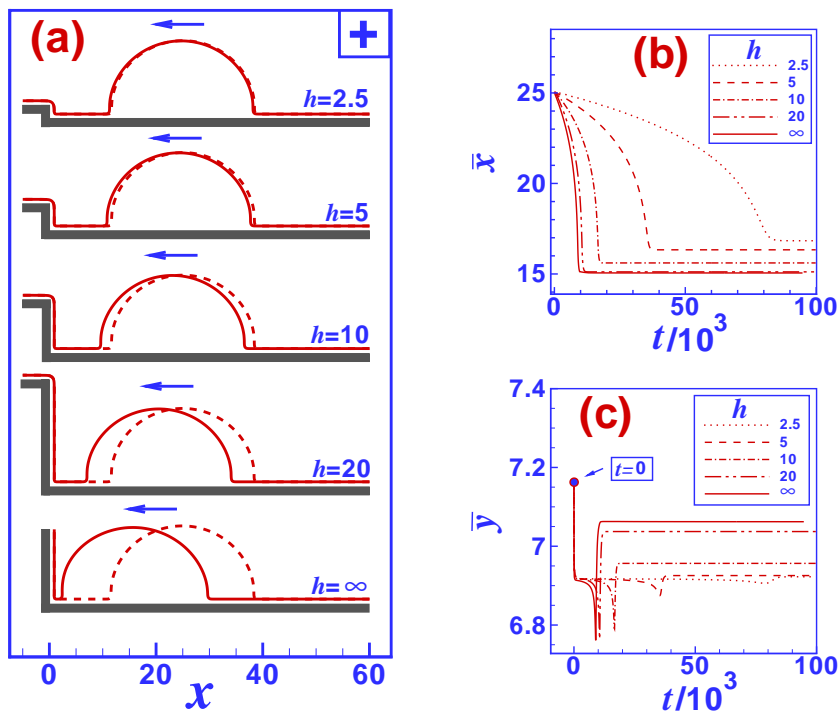}
\caption{(a) Droplets of initial size $a=15$ on substrates with $\theq=90^\circ$
($C=4.2327$ and $B=-2.5$) starting at $\ell=10$ on the base of
steps with heights varying between
$h=2.5$ (top) and $h=\infty$ (bottom, corresponding to an
isolated wedge) right after the initial relaxation process at
$t=210$ (dashed lines) and during the migration process at $t=9100$ (solid lines)\/. 
The horizontal position $\bar{x}$ and the vertical position
$\bar{y}$ of the center of mass are shown in (b) and (c),
respectively, as a function of time.} 
\label{wedgediffhplus}
\end{figure}
%=======================================================================
%==========================   FIGURE  ==================================
\begin{figure}[b]
\includegraphics[width=0.8\linewidth]{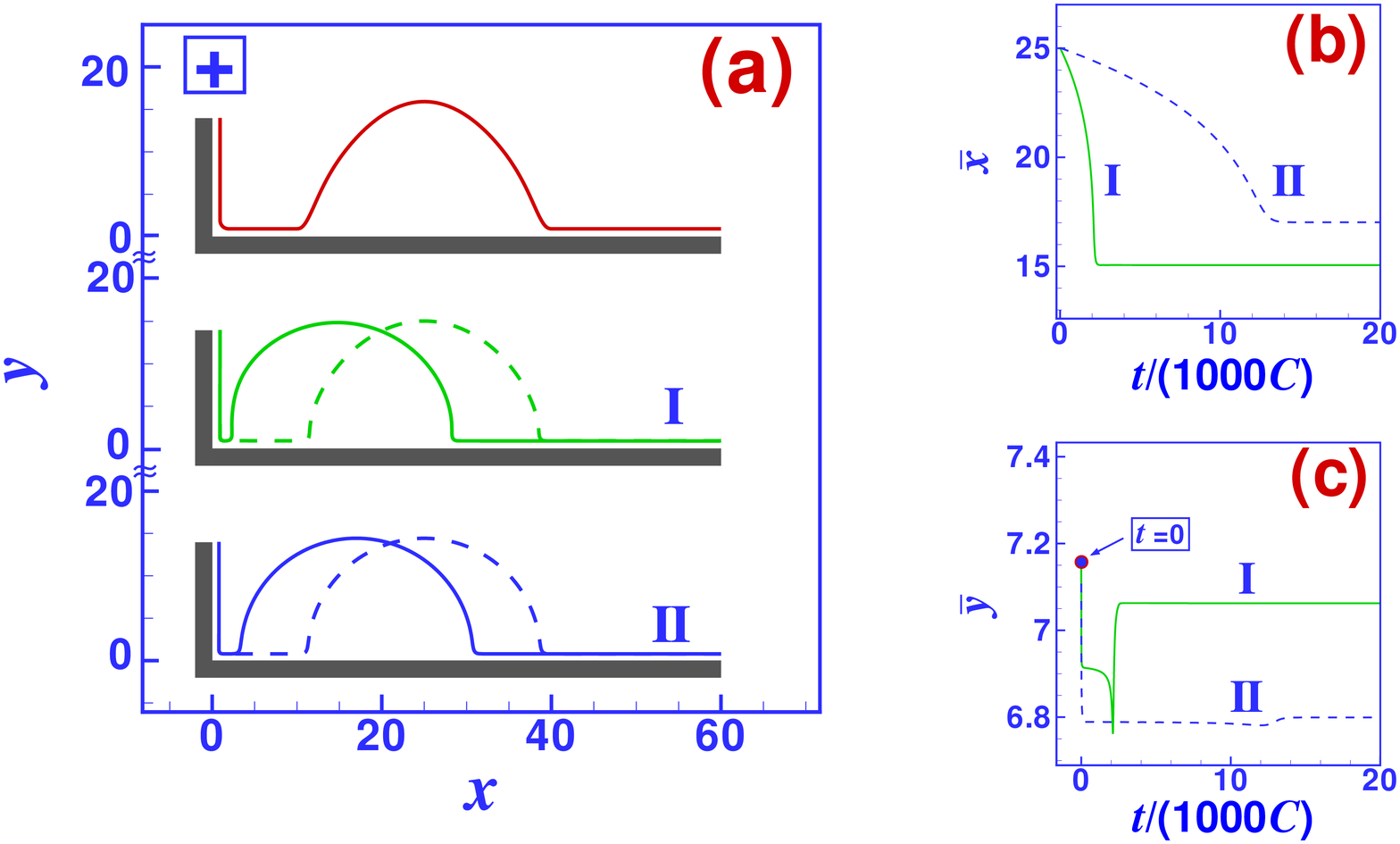}
\caption{Droplets of initial size $a=15$ near isolated wedges
($h=\infty$) in the plus
case. $\theq=90^\circ$ on both substrates: (I: $B=-2.5$, $C=4.2327$)
and (II: $B=-4$, $C=0.9265$)\/. (a) The top panel shows the initial
droplet shape and the lower panels show the droplets just after the initial
relaxation at $t/C\approx 60$ (dashed lines)  and in their final 
configuration at $t/C=2500$ and $t/C=14500$ for substrate I and II,
respectively (solid lines)\/. The horizontal position $\bar{x}$
and the vertical position $\bar{y}$ of the center of the mass during the motion
are shown in (b) and (c), respectively.
%In order to compare the profile the times are scaled by $C$ values.
%\cite{moosavi08a}.}
\label{wedgediffbplus}}
\end{figure} 
%=======================================================================

In the plus case the direction of motion of the droplets is
reversed as compared to the minus case. As shown in
Fig.~\ref{wedgediffhplus}(a), the
migration speed increases with the step height, but the droplets stop
before the leading contact line reaches the wedge such that the
droplets do not move into the corner. As in the other cases
discussed so far, the droplet speed increases significantly as
the step height is increased up to $h=20$\/. In
Fig.~\ref{wedgediffhplus}(b) the trajectories $\bar{x}(t)$ for $h=20$ and
for $h=\infty$ almost coincide. The final distance of the droplets
from the wedge decreases with the step height, but it remains
finite in the limit $h\to\infty$\/. Once the droplets reach the
wedge there is a brief drop of $\bar{y}$ due to fluid
leaking out of the droplet into the corner area\/. After that their
vertical position $\bar{y}$ of the center of mass increases
again (see Fig.~\ref{wedgediffhplus}(c))\/. This increase is the result
of a contraction of the droplets, which is also observed for
droplets on the top side of steps in the minus case and which is more
pronounced for higher steps.

Changing $B$ and $C$ such that $\theq=90^\circ$ remains the same 
does not change the dynamics of the droplets qualitatively. Higher
values of $B$ and $C$ result in larger droplet velocities
(see Fig.~\ref{wedgediffbplus} concerning the example of droplets near an
isolated wedge)\/.
Beside this change of droplet speed we find that for larger values of $C$
the final position of the droplets is closer to the step and the
final height $\bar{y}$ of their center of mass is larger. The
magnitude of the disjoining pressure and
therefore the forces acting on the droplet increase with $C$\/. In
response to these forces the droplets deform more upon increasing
$C$\/.

%=========================== SUBSECTION ================================
\subsection{Nanodroplets on edges, wedges, and steps}

In the previous subsections we have discussed the behavior of
nanodroplets originally positioned at a certain lateral distance from
topographic features such as edges, wedges, and steps. 
Their behavior suggests that these surface features provide
migration barriers for droplets. Even
in those situations in which droplets migrate towards the edge or
wedge, respectively, they stop just before reaching them.
This result is also borne out in a macroscopic model which takes into account
only interface energies: the free energy of a droplet positioned
right on an edge is larger than that of a droplet of equal volume 
residing on a flat and homogeneous substrate, and the free energy
of a drop in the corner of a wedge is even lower. As a consequence,
we expect that 
droplets sitting on edges to be in an unstable and droplets sitting
inside the corner of a wedge to be in a stable configuration.
Moving, by force, a droplet (with the shape of the liquid-vapor interface remaining a part of a
circle) in the first case slightly to one side results in an
increased contact angle on this side, while the contact angle on
the other side decreases. However, with the leading contact angle
being larger than the equilibrium one the corresponding contact
line will move away from the edge, while the trailing contact line
(with the corresponding contact angle being smaller than the
equilibrium one) moves towards the edge. As a consequence, the
droplet leaves its position at the edge. In the case of a droplet
in a wedge, the situation is reversed: moving, by force, the droplet 
into one
direction results in a decreased contact angle on this side and an
increased contact angle at the trailing side, such that the
droplet moves back into the corner of the wedge.
Accordingly one expects that a certain force has to be applied
to push a droplet over an edge or to pull it out of the
corner of a wedge. 
In the following our detailed numerical results indicate that this
also holds for nanodroplets and they enable us to quantify those
external forces.

\begin{figure}
\includegraphics[width=0.8\linewidth]{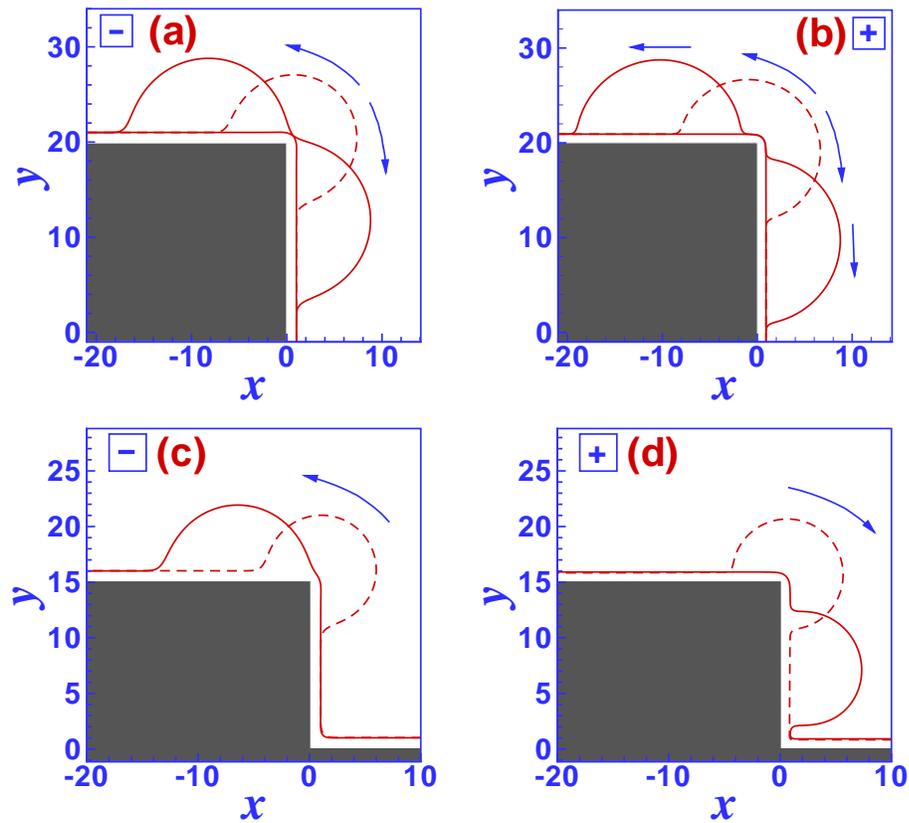}
\caption{A droplet positioned symmetrically on an
edge (dashed line) is in an unstable equilibrium both in the (a) minus
case ($C=2.67$ and $B=0$) and in the (b) plus case ($C=4.2327$ and
$B=-2.5$)\/. A tiny perturbation at $t=0$ pushes the droplet
either up or down the step. After that in the minus case the droplet stays
next to the step (a) whereas in the plus case (b) it moves away
from the edge (solid lines, $t=1000$ and $t=1050$ in (a) and (b),
respectively)\/. In the presence of a wedge, i.e.,
for a finite step height $h$,  the droplet is (c) pushed
onto the top side of the step in the minus case but (d) towards the
corner of the step in the plus case (solid lines, $t=800$ and
$t=815$ in (c) and (d), respectively)\/.}
\label{overedge}
\end{figure}

Our analyses show that a nanodroplet positioned
symmetrically on the tip of an edge is unstable on all types of
substrates, regardless of whether the droplets migrate towards the
edge or away from the edge (see Fig.~\ref{overedge}; there a
suitable, highly
symmetric initial shape of the liquid-vapor interface
has been chosen such that it indeed relaxes to the unstable state of 
a droplet sitting on the tip of the edge)\/.
Due to the mirror symmetry with respect to the diagonal of the edge, a droplet
right at the tip of an isolated edge is in mechanical
equilibrium but in an unstable one. In the minus case, after a small perturbation the
droplet flips either up or down but  then rests next to the step,
i.e., in the position which it would assumes upon migrating towards the
edge (Fig.~\ref{overedge}(a))\/. In the plus case, as expected from
the previous results the droplet
migrates away from the edge after flipping to either side
(Fig.~\ref{overedge}(b))\/. At  steps of finite height,
this symmetry is broken by the presence of the wedge.
In the minus case, the droplets are
pushed away from the wedge, i.e., upwards, which is consistent
with the dynamics of droplets in the vicinity of isolated wedges of
the same material. However, being attracted to the edge as shown 
in the previous subsections, they come
to rest with the trailing contact line pinned to the step edge
(Fig.~\ref{overedge}(c))\/.
In the plus case the droplets move in the opposite direction, i.e.,
they are attracted by the wedge which they migrate to after leaving
the edge area (Fig.~\ref{overedge}(d))\/.

\begin{figure}
\includegraphics[width=0.8\linewidth]{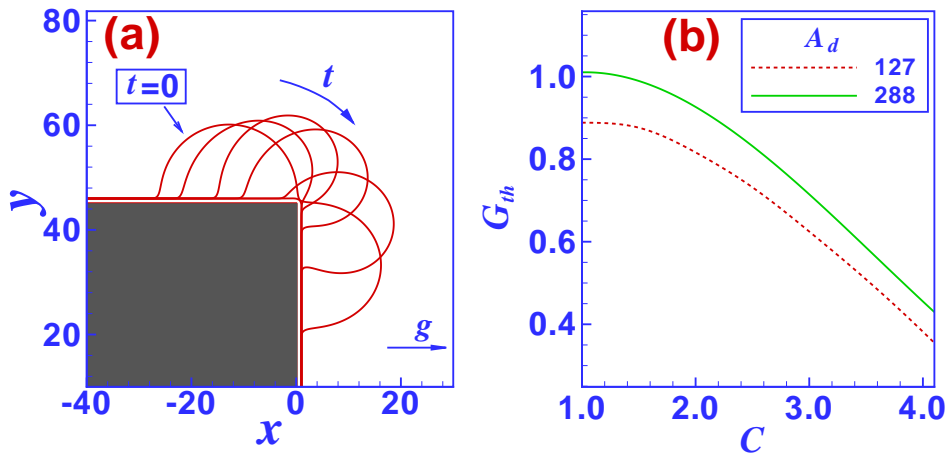}
\caption{ (a) A nanodroplet of radius $a=15$ pushed over an edge
(minus case, $B=0$ and $C=3$) by an external,
horizontal  body force $g\,\vct{e}_x= 0.00208\,\vct{e}_x$
($G=g\,A_d=0.71475$, direction indicated by the
horizontal arrow)\/. Droplet shapes for $t=0$ (indicated), 100, 650, 1050,
1150, and 6950 are shown (from the upper left to the lower
right)\/.
(b) The minimum total force per unit length $G_{th}=g_{th}\,A_d$ ($A_d$ is the 
droplet cross-sectional area) to push droplets over the edge
for two droplet cross-sectional areas
$A_d=127$ (dashed line) and
$288$ (solid line) corresponding to $a\simeq10$ and 15,
respectively. The values of $(C,B)= (1,0)$, $(2,0)$, $(3,0)$, and
$(4,0)$, 
correspond to  $\theq=51.3^\circ$ , $75.5^\circ$, 
$97.2^\circ$, and $120^\circ$, respectively. The force density
(force per unit volume) $g$
is measured in units of $\gamma/{b^2}$.}
\label{edgeforce}
\end{figure}

In order to displace a droplet from one side of an edge to the 
other side (as shown in Fig.~\ref{edgeforce}(a)) one has to apply an
external force, e.g., a body force such as gravity or centrifugal
forces, which we incorporate into the hydrodynamic equations via the
boundary condition (see Eq.~\eqref{eq:surfacebc})\/. 
If the applied force is small the droplets assume a new but
distorted equilibrium position with the leading three-phase contact
line still pinned at the edge. But there exists a threshold force
density $g_{th}$
beyond which the configuration described above is unstable and the
leading three-phase contact line depins from the step edge. As
a consequence the droplet flips around the corner and ends up on
the vertical side of the edge. Since the applied body force has no
component parallel to this vertical part of the substrate, the further fate
of the droplet is determined by the action of the intermolecular
forces. In the minus case considered in Fig.~\ref{edgeforce}(a) the
droplet is attracted to the edge such that the
new stable equilibrium configuration is that of a droplet residing
on the vertical part of the step with the trailing three-phase
contact line pinned at the step edge. In the plus case (which we
have not tested numerically) the droplet is repelled from the edge
and it is expected to move down the vertical part of the edge.
As shown in Fig.~\ref{edgeforce}(b) we have determined
the body force density $g$ needed to push the droplets
over the edge for various types of substrates (minus case with
$B=0$) and for droplets of two different sizes.
The threshold force density $g_{th}$ decreases both with $C$
(i.e., with $\theq$) and with the droplet size. Both trends are also
expected to occur for macroscopic droplets. In the limit $\theq\to
180^\circ$ the droplets loose contact with the substrate and the
free energy of the droplet at the edge equals the free energy
on a planar substrate. Taking, however,  the finite range of
molecular interactions into account this no longer holds, but the barrier still
decreases with increasing $\theq$\/. Since the force density $g$ is a body
force, i.e., a force density, the total force per unit ridge length
$G=g\,A_d$ (with the ridge cross-sectional area $A_d=\Omega_d/L\sim
a^2$) acting on the droplet
is proportional to the droplet cross-sectional area $A_d$\/.
In Fig.~\ref{edgeforce}(b) we observe that the total threshold
force $G_{th}$ needed to push droplets over the edge increases with
the droplet volume.  Apart from the effects of
long-ranged intermolecular forces the main contribution to the
barrier effect of the edge is the increase of the liquid-vapor surface
area when the droplet is deformed as it passes over the edge.
The square root of the ratio of the surface tension coefficient
$\gamma$ and the body force density $g$ defines a capillary length below which
the surface tension dominates, while it is less important for 
larger drops. (The surface area of three-dimensional droplets
increases only quadratically with the droplet
radius $a$ while the volume increases $\sim a^3$\/.) 
From dimensional arguments $g_{th}\,a^3\sim \gamma\,a^2$ one
expects the threshold body force density $g_{th}$ 
needed to push droplets over an edge to decrease $\sim 1/a$
with the droplet radius,
while for liquid ridges the total force per unit length $G_{th}$
should still increase linearly with the droplet radius $a$\/.
Therefore the total threshold force $G_{th}$ needed for the larger droplet in
Fig.~\ref{edgeforce}(b) should be about $\sqrt{288/127}\approx 1.5$
times the force needed for the smaller drop. The actual value is somewhat
smaller and we attribute the difference to the effect of the
long-ranged part of the intermolecular forces.

%==========================   FIGURE  ==================================
%=======================================================================
%==========================   FIGURE  ==================================
%=======================================================================
%==========================   FIGURE  ==================================
\begin{figure}
\includegraphics[width=0.8\linewidth]{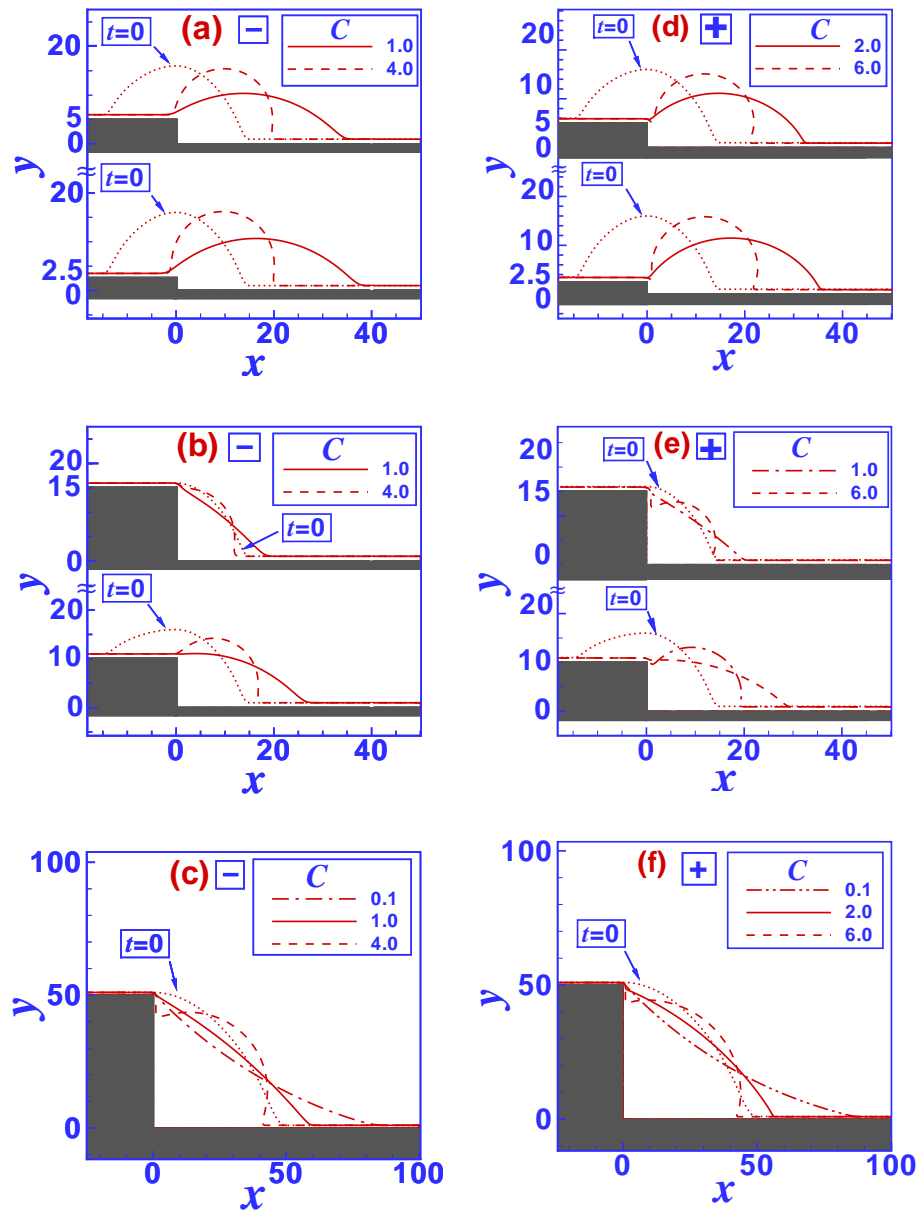}
\caption{The final positions and shapes of droplets initially
(dashed lines) spanning 
a topographic step of height $h=5$ ($A_d=250$) and $h=2.5$
($A_d=275$) ((a) and (d)), $h=15$ ($A_d=150$) and
$h=10$ ($A_d=200$) ((b) and (e)), and  $h=50$ ($A_d=1667$, (c) and (f)) in 
the minus case ($B=0$, (a), (b), and (c))
and in the plus case ($B=-2.5$, (d), (e), and (f)), for various values
of $C$\/.  
In the minus case $C=0.1$, $1$, 
and $4$ correspond to $\theq=15.7^\circ$, $51.3^\circ$, and $120^\circ$, 
respectively, while for the plus case $C=0.1$, $1$, $2$, and $6$ correspond to 
$\theq=12.5^\circ$, $40.2^\circ$, $58.2^\circ$, and $114.7^\circ$, 
respectively.}
%In contrast to the case of droplets near the step, here the droplets have not been displaced away from the step.
\label{overwedge}
\end{figure}
%=======================================================================
%====%==========================   FIGURE  ==================================
\begin{figure}
\includegraphics[width=0.8\linewidth]{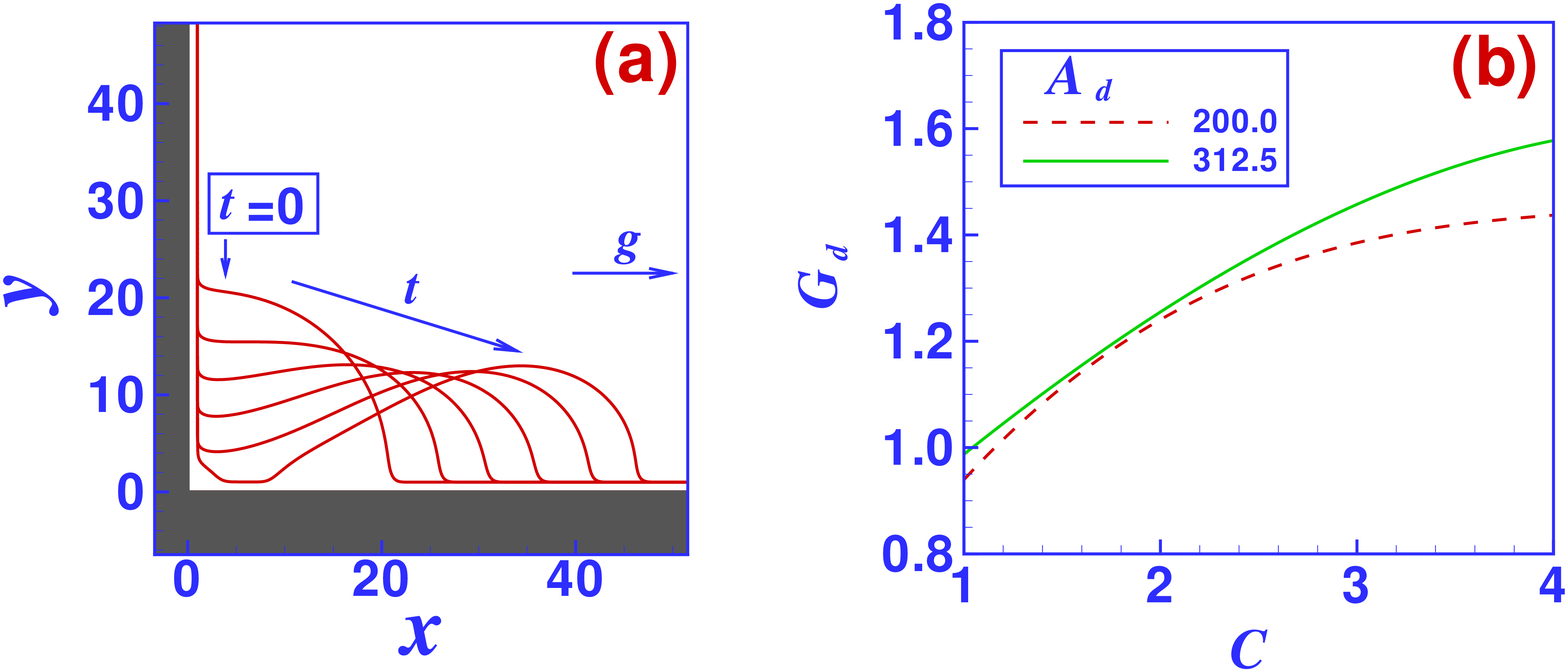}
\caption{ (a) A droplet with cross-sectional area $A_d=312.5$
pulled out of the corner of a wedge ($B=0$ and $C=3$) by an
external horizontal body
force $g\,\vct{e}_x=0.00465\,\vct{e}_x$ ($G=g\,A_d=1.453125$, direction indicated by the
horizontal arrow)\/. Shown are droplet shapes for $t=0$ (indicated),
400, 1400, 3700, 4700, and 4900 (from left to right)\/.
(b) The minimum total force per unit ridge length $G_{th}= g_{th}\,A_d$ required to extract the
droplet from the corner as a function of
$C$ for droplets of cross-sectional area  $A_d=200$ (corresponding to
$a\simeq17.5$, dashed line) and $A_d=312.5$ (corresponding to
$a\simeq22$, solid line) in the minus case ($B=0$)\/. 
The values $C=1$, $2$, $3$, and $4$ correspond to
$\theq=51.3^\circ$ , $75.5^\circ$, $97.2^\circ$, and $120^\circ$,
respectively. The force density $g$ is measured in units of $\gamma/{b^2}$.}
\label{wedgeforce}
\end{figure}
%=======================================================================
%Increasing the volume slightly 
%increases the required applied force to to displace the droplets on the edge, which can be 
%associated to the further DJP imposed on the larger droplet. As a result, the 
%required applied force density (acceleration) considerably decreases
%because of its proportionality with the volume of the dropelts. 

%the applied force on the droplets is 
%proportional to the volume of the droplets while the resistance forces do not 
%increase at the same level.

A macroscopic droplet spanning the whole topographic step (i.e., with one
contact line on the top terrace and one on the base terrace) moves
downhill: since the surface of a macroscopic droplet is a part of a
circle which is cut by the top side of the step at a higher level
than by the base
of the step the contact angle at the upper terrace is
smaller than the contact angle at the lower terrace. This results
in a net driving force in downhill direction. The final
configuration is a droplet with the upper contact line pinned at
the step edge. This is also true on the nano scale, as demonstrated
in Figs.~\ref{overwedge}(a) and \ref{overwedge}(b) for the minus case
and in Figs.~\ref{overwedge}(d) and \ref{overwedge}(e) for the plus
case. The latter indicates that the difference in contact angle at
the two contact lines due to the different height level at the top
side and at the base side of the step provides a stronger driving force 
than the lateral action of the disjoining pressure which, in the plus case,
moves droplets positioned next to the step in uphill direction. 

For all substrates the surface of droplets pinned at the edge
becomes convex (corresponding to a negative pressure in the
droplet) for small $\theq$ (i.e., small $C$, see $C=0.1$ in
Figs.~\ref{overwedge}(c) and \ref{overwedge}(f))\/. For very large
$\theq$ the upper contact line depins from the edge and moves down
towards the wedge (see $C=6$ in Figs.~\ref{overwedge}(c) and
\ref{overwedge}(f))\/. The result is a droplet sitting in the
corner of the wedge
area only. The critical value for $\theq$ between both types of
configurations depends on the
droplet volume and the step height: the smaller the droplet (as
compared to the step height) the smaller is the value of $\theq$ at
which the upper contact line depins and the larger the volume the
smaller is the value of $\theq$ at which the droplet surface becomes convex.
Both phenomena are in qualitative
agreement with macroscopic considerations which take into account
interface energies only (see Refs.~\cite{seemann05,khare07})\/.

Droplets sitting in the corner of a wedge are in an energetically rather
favorable situation as illustrated by the arguments given at the
beginning of this subsection. 
However, even in the plus case, for which droplets are
attracted by wedges, they stop before reaching the wedge and they do
not move into the corner of the wedge. In any case, there is an energy barrier
to overcome in order to move droplets out of wedges, as
shown in Fig.~\ref{wedgeforce}(a)\/. 
If a small horizontal force is applied to a droplet sitting in
the corner of a wedge it assumes a new, slightly distorted but
stable shape. But there exists a threshold force density $g_{th}$
above which the distorted configuration becomes unstable and the
droplet moves out of the corner. In the minus case considered in
Fig.~\ref{wedgeforce}, the droplet is repelled from the wedge such
that the effect of the intermolecular forces adds to the external
driving force and the droplet definitively moves out of the corner of the
wedge. In contrast to the force required to push droplets over an
edge, $g_{th}$ increases with $C$
(i.e., with $\theq$)\/. The total force per unit ridge length
$G_{th}=g_{th}\,A_d$ required to pull a droplet
out of a wedge increases only slightly (i.e., less than linearly)
with the droplet volume (see Fig.~\ref{wedgeforce}(b)), so that accordingly
the required force density $g_{th}$ decreases significantly with
volume. With the same dimensional arguments used
above for droplets being pushed over edges one would expect the
total force needed to pull two droplets of different volume 
out of the corner of a wedge to
be proportional to the square root of the volume ratio. In particular for
small $C\approx 2$ according to Fig.~\ref{wedgeforce}(b) the total
threshold force is almost independent of the droplet size, rather
than to increase by a factor $\sqrt{312.5/200}=1.25$\/. We attribute this
difference to the influence of the long-ranged part of the intermolecular
forces.

%%%%%%%%%%%%%%%%%%%%%%%%%%%%%% SECTION %%%%%%%%%%%%%%%%%%%%%%%%%%%%%%%%%
\section{Discussion}
\label{discuss}
%=========================== SUBSECTION ================================
\subsection{Force analysis}
\label{force}
Numerical solutions of the Stokes dynamics of nanodroplets in the
vicinity of edges, wedges, and steps are rather time consuming,
even when using advanced numerical methods. As shown in
Fig.~\ref{comppisigma}(b), the main driving force for the migration
of droplets is the disjoining pressure induced force density $f_\Pi$ as
defined in Eq.~\eqref{dpforce}\/. After the initial relaxation
process, the shapes of the droplets hardly change during the
migration process until the droplets either reach the
edge (minus case) or the corner of the wedge area (plus case)\/. 
Unfortunately the relaxed shape of the droplet is not available
analytically, but for droplets on substrates with $\theq\approx
90^\circ$ as mostly considered here the initial shape relaxations
are rather mild. Accordingly, as demonstrated in the following, the
force on the droplets can be estimated rather accurately
from calculating $f_\Pi$ for droplets with a shape given by the
initial profile in Eq.~\eqref{inicond} positioned at the 
distance $\ell=|\bar{x}|-a$ from the edge or from the corner of the wedge\/. 
Apparently this estimate
becomes invalid for $\ell\lesssim 1$\/.

\begin{figure}
\includegraphics[width=0.8\linewidth]{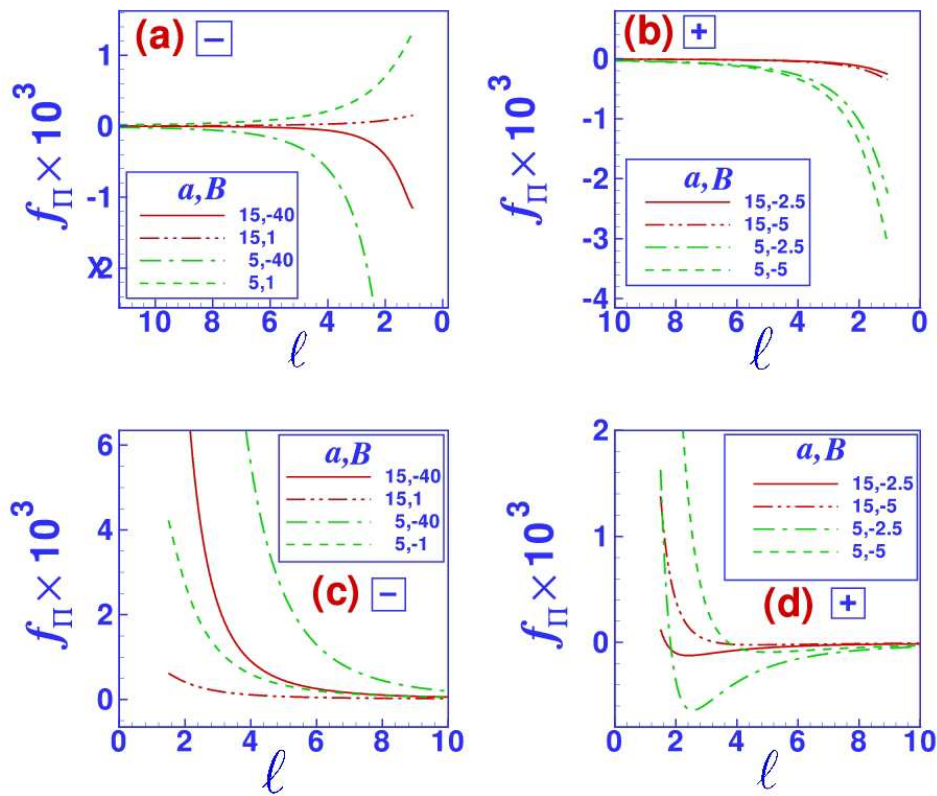}
\caption{The DJP induced force density $f_\Pi$ (in units of $\gamma/b^2$) 
on droplets of size $a=15$ and $a=5$
in the vicinity of an edge ((a) and (b)) and a wedge ((c) and (d)) on
substrates of the minus ((a) and (c)) and the plus ((b) and (d)) type with
$C=1$ and various values of $B$ as indicated in the boxes as a
function of the distance $\ell$ from the edge or the corner of the wedge. 
\label{dropsize} }
\end{figure}

Figures~\ref{dropsize}(a) and \ref{dropsize}(b) show the disjoining
pressure induced force densities calculated along these lines for
droplets of size $a=15$ and $5$ as a function of the distance $\ell$ of the
right contact line to an edge for the minus and the plus case,
respectively. Since $f_\Pi$ is proportional to $C$ only results for
$C=1$ are shown. 
For the 
plus case the force is always negative for both droplet sizes and
at all distances from the edge, with
its strength increasing towards the edge. This means that 
droplets should move away from the edge with a speed which decreases
continuously. This is in complete agreement with
the numerical results presented in the previous section. 
For the minus case and for sufficiently large values of $B$, the
force is positive in accordance with the numerical results. However,
as shown in Fig.~\ref{dropsize}(a), for very small values of
$B$, i.e., for $B<B_c\simeq-10$ the force in the direct vicinity of the edge
becomes negative and droplets are expected to move away from the
edge. Indeed, as demonstrated in Fig.~\ref{edgediffb}(b)
($\bar{x}(t\to\infty)$ for I lies below the corresponding values
for II and III), the final
distance of the droplets from the step edge in the minus case
increases with more negative values of $B$\/. On the other hand, as
shown in the following Subsec.~\ref{direction}, for large
distances from the edge, in the minus case even for arbitrarily
small $B$ the force is positive so that droplets find an
equilibrium position with vanishing force at a
significant distance from the edge. The sign of the disjoining
pressure induced force density does not
depend on the droplet size. However, the equilibrium position
changes as a function of droplet size.
%%%reveals that besides the strength, sign of the force may also
%%%change for small values of $B$.  For $B>B_c\approx-10$ the force is
%%%always positive which means that the droplets move towards the
%%%edge. But for $B<B_c$ the force changes sign at $\ell=\ell_c>1$
%%%from the positive to negative at $\ell=\ell_c$. Thus if the
%%%droplets are positioned at $\ell<\ell_c$ $(\ell<\ell_c )$ they move
%%%away (towards) the edge and stop at $\ell=\ell_c$. $\ell_c$
%%%increases with decreasing (increasing) the value of $B$ (size of
%%%the droplets $a$). The results of the performed simulations
%%%%for a larger negative $B$ ($B=-40$) 
%%%verifies that this is indeed 
%%%correct and a droplet moves away from the edge for very negative values of $B$ if it is positioned at distances smaller than $\ell_c$. 
%(see Fig.~\ref{edgem40}). 

The force calculated for droplets of the same size but in the
vicinity of a wedge for the minus and the plus case are 
shown in Figs.~\ref{dropsize}(c) and \ref{dropsize}(d), respectively. For
the minus case the force is positive for any droplet size and for
any $B$,
which means that the droplets move away from the wedge. For the
plus case the force is negative at large distances, but it changes
sign close to the wedge at a distance which increases with
decreasing the size of the droplets and with decreasing the value
of $B$\/. The latter relation
is in agreement with the numerical results presented in
Fig.~\ref{wedgediffbplus}(b)\/.
%%%Therefore the droplets should move
%%%towards the wedge when it positioned far enough away from the
%%%wedge. The droplet stops at a distance from the wedge which
%%%decreases (increases) with decreasing $B$ (size of the droplets
%%%$a$). 

\begin{figure}
\includegraphics[width=0.8\linewidth]{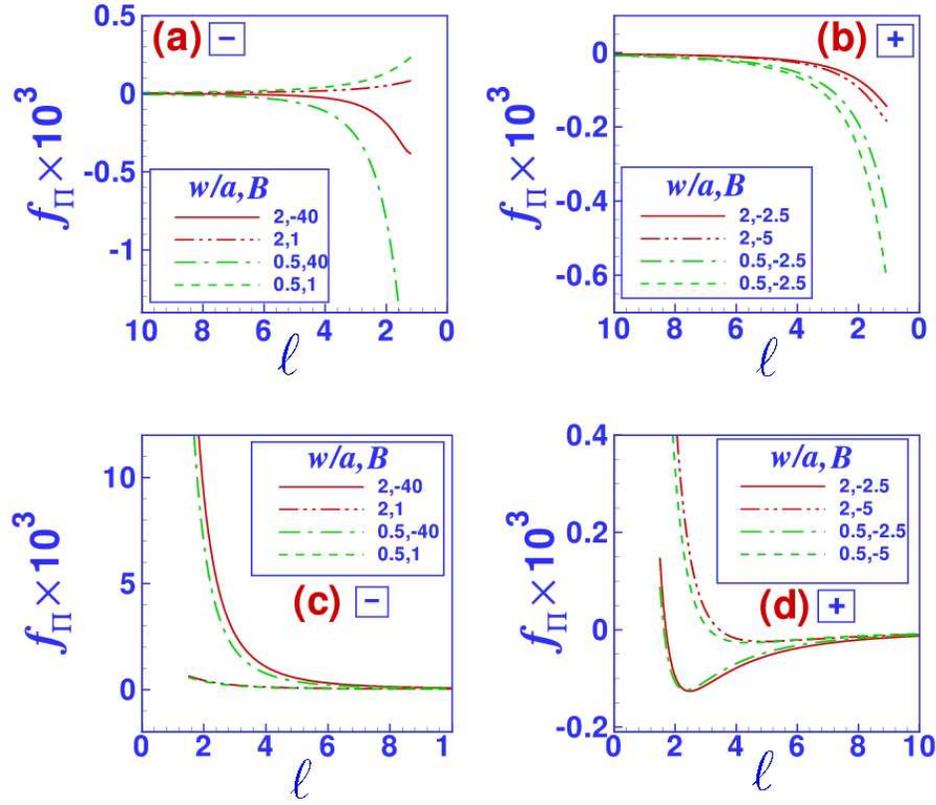}
\caption{The DJP induced force density $f_\Pi$ (in units of $\gamma/b^2$) 
on droplets of height $a=15$
and widths $w=2\,a$ and $w=0.5\,a$ in the vicinity of an edge ((a)
and (b)) and a wedge ((c) and (d)) on substrates of the minus ((a)
and (c)) and the plus ((b) and (d)) type with $C=1$ and various values of $B$
as indicated in the boxes as a function of the distance $\ell$
from the edge or wedge.  \label{profsize}}
\end{figure}

The disjoining pressure induced force density $f_\Pi$ presented in
Fig.~\ref{dropsize} has beeen calculated for droplets with a shape given
by Eq.~\eqref{inicond}, i.e., for droplets with
equal height and half width. However, the substrate parameters used
in Fig.~\ref{dropsize} do not necessarily lead to $\theq=90^\circ$,
and droplets would adopt a very different shape even during the
migration process. In order to check the influence of the droplet
shape on the calculated disjoining pressure induced force density $f_\Pi$
we also consider droplets which have a
width $w$ different from their height $a$ (compare with
Eq.~\eqref{inicond}):
\begin{equation}
y(x)= y_0+
a\,\left[1-\left(\frac{{|x-\bar{x}|}}{w}\right)^2\,\right]^{|x-\bar{x}|^m+1}.
\end{equation}
%%%The calculation of the force so far is only performed on droplets with their heights equal to their base widths. However, based on the equilibrium contact angle of the system and size of the droplets, the droplets can have different configurations; one should check whether considering different configurations can change the results. The effect of the configuration of the droplets on the calculated force is shown 
%%%in Figure~\ref{profsize}. In the calculation of the force it is assumed 
%%%that the width of the droplet can be different from the height of the 
%%%droplet. Thus, a profile of the form 
%%%\begin{equation}
%%%y(x;t=0)= y_0+a\,\left[1-\left(\frac{{|x-\bar{x}|}}{w}\right)^2\,\right]^{|x-\bar{x}|^m+1}
%%%\end{equation}
%%%with $w$ as the half base width of the droplet is considered for the force calculation. 
Figure~\ref{profsize} compares the disjoining pressure induced
force density $f_\Pi$ on droplets in the vicinity of edges and wedges in both the
minus and the plus case for different drop widths $w$ but for a
fixed drop height $a=15$\/. The results
%, in comparison with those of Fig.~\ref{dropsize}, 
indicate that the form of the droplets does not change the sign of
$f_\Pi$, and in particular in the vicinity of the wedge the droplet
width has a rather small influence on the force.

\begin{figure}
\includegraphics[width=0.8\linewidth]{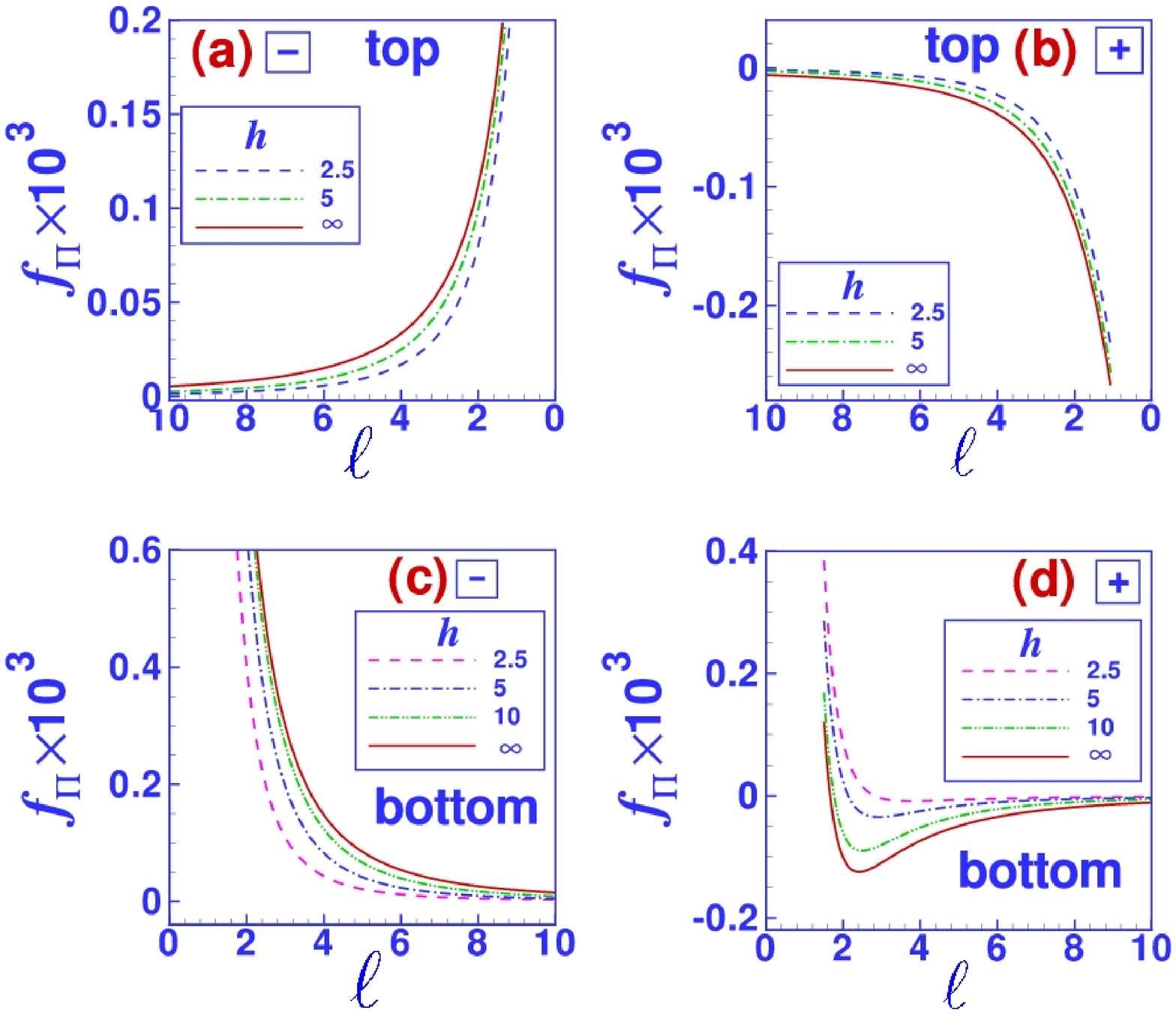}
\caption{The DJP induced force density $f_\Pi$ (in units of
$\gamma/b^2$) 
on droplets of size $a=w=15$
on the top side ((a) and (b)) and on the bottom side ((c) and (d))
of steps of various heights $h$ with 
substrates of the minus ($B=-1$, (a) and (c)) and the plus
($B=-2.5$, (b) and (d)) type with $C=1$ as a function of the
distance $\ell$ from the step\/.
\label{stepsize} }
\end{figure}

In the vicinity of topographic steps the dependence of $f_\Pi$ on
the step height $h$ is also in good agreement with the results of the
full numerical solution of the Stokes dynamics. 
Figure~\ref{stepsize} shows $f_\Pi$ above and below the step on
substrates of the minus and plus type for step heights ranging from
$h=2.5$ to $\infty$ (i.e., to isolated edges and wedges)\/. The
absolute value of the force increases with the step height with the
force near isolated edges and wedges as the limiting values.
This limiting value is almost reached for a  step height
$h=20$ (not shown in Fig.~\ref{stepsize})\/.

%==========================   FIGURE  ==================================
%=======================================================================
%==========================   FIGURE  ==================================
%\begin{figure}
%\includegraphics[width=0.45\linewidth]{edgem40}
%\caption{ Motion of a nanodroplet with $a=5$ near an edge for the 
%minus case ($C=0.01275$, $B=-40$: $\theq=90^\circ$). 
%The droplet moves away from the edge if it is positioned very 
%close to the edge. The dashed and the solid lines correspond 
%to $t=0.46$ and 601, respectively.}
%\label{edgem40}
%\end{figure}
%=======================================================================
%==========================   FIGURE  ==================================
%=======================================================================
%==========================   FIGURE  ==================================
%==========================================================================================================================================

%=========================== SUBSECTION ================================

\subsection{Direction of motion far from the step}
\label{direction}

Both the force calculations presented in the previous subsection as
well as the results of the numerical solution of the mesoscopic
hydrodynamic equations indicate, that the 
direction of motion of a nanodroplet far enough from the 
step does not depend on whether the droplet is positioned on the
top side or on the bottom side of 
the step. In the minus case the droplets move in  downhill direction 
(i.e., in the direction of positive $x$-values) and in the plus case in the
opposite direction, independent of the step height and of the values
of $B$ and $C$ (where the latter has to be positive)\/. 
In order to understand this we further analyze the total force per
unit ridge length $F_\Pi=f_\Pi\,A_d$ on liquid ridges
as defined in Eq.~\eqref{dpforce} for large droplets far
from the step\/. 
Asymptotically for large $|x|$ the DJP reduces to its value on a
flat substrate so that there the wetting film thickness assumes its
value $y_0$ independent of $x$ up to $O(|x|^{-3})$\/.
The leading order correction to the DJP is $\pm \sign (x)\,
9\,h\,C/(16\,x^4)$ for the plus and minus case, respectively.
Parameterizing the shape of the liquid-vapor interface to the left and to the
right of the droplet apex by $x_\ell(y)$ and $x_r(y)$,
respectively, from Eq.~\eqref{dpforce} with $dy=-n_x\,ds$ we obtain:
\begin{eqnarray}
F_\Pi&=&-\left[\int_{y_0}^{y_m} \Pi(x_\ell(y),y)\,dy
-\int_{y_0}^{y_m} \Pi(x_r(y),y)\,dy\right]\nonumber\\
&\approx& \pm \sign(x)\,\int_{y_0}^{y_m} \frac{9\,h\,C}{16}\,\left[
\frac{1}{x_r(y)^4}-\frac{1}{x_\ell(y)^4}
\right]\,dy\nonumber\\
&\approx& \mp \frac{9\,h\,C}{16\,|\bar{x}|^5}\,\int_{y_0}^{y_m}\left[
x_r(y)-x_\ell(y)\right]\,dy=\mp
\frac{9\,h\,C}{16\,|\bar{x}|^5}\,A_d\/,
\label{flimit0}
\end{eqnarray}
with the droplet apex height $y_m$\/. In the last but one step we
have approximated $F(x_r)-F(x_l)\approx (x_r-x_l)\,F'(\bar{x})$ with
$F(x)=x^{-4}$\/.
The force is proportional to the droplet cross-sectional area and its sign is
determined by the sign of the Hamaker constant (i.e., depending on
the case; plus or minus) only: in the plus case the force is negative
(upper sign) and in the minus case it is positive (lower sign)\/.
This is in complete agreement with the
numerical data. However, other than suggested by
Eq.~\eqref{flimit0}, the force on a droplet 
does not diverge in the limit $h\to\infty$ as this limit has to be
taken before taking the limit $|x|\to\infty$\/. 

At large distances from an isolated wedge as well as from an
isolated edge, the disjoining pressure is to leading order given by
the DJP of the corresponding homogeneous substrate with
$\pm\,C/(2\,x^3)$ as the leading order correction for the plus and
the minus case, respectively. As in the case of the step, up to
this order the thickness of the wetting film is independent of the
(large) distances from the edge or wedge. 
Using the same approximations as in the case of the step of finite
height, the force on a droplet at a distance $|\bar{x}|$ from an
edge is given by
\begin{equation}
F_\Pi=\mp\,\frac{3\,C}{2\,\bar{x}^4}\,A_d,
\end{equation}
with the upper sign corresponding to the plus case and the lower
sign to the minus case.
The sign of the force is the same as in the case of a step and it
is also proportional to the droplet volume. However, it decreases
less rapidly with the distance from the step.

For very large, almost macroscopic droplets, the situation is again
different from the previous two. In the following we follow the
line of arguments developed in Refs.~\cite{moosavi08a,moosavi08b}
for droplets in the vicinity of chemical steps.
In this limit the droplets are approximately symmetric with respect
to their apex and the main contribution to the
force stems from the vicinity of the contact lines. For the wetting
film as well as near the apex the $x$-component $n_x$ of the surface
normal vector is zero and thus in the vicinity of the apex the DJP
is negligibly small. In most of the examples discussed here  the
equilibrium contact angle $\theq$ is about $90^\circ$ and, as a
consequence, the lateral width of the contact lines (i.e., the range
of $x$-values within which the drop profile crosses over to the
flat one of the wetting film) is small and the
lateral variation of the DJP within this region is negligible.
Therefore, after parameterizing the droplet surface in the vicinity
of the left and right contact line (at $x=\bar{x}-a$ and
$x=\bar{x}+a$, respectively) by the corresponding function $x(y)$,
the total force on a droplet can be approximated by 
\begin {eqnarray} 
\nonumber
 F_\Pi&\approx&-\Bigg[\int_{y_0}^{\infty}\Pi(\bar{x}-a,y)\,dy
 -\int_{y_0}^{\infty}\Pi(\bar{x}+a,y)\,dy\Bigg] \\ 
&=& -\Phi(\bar{x}-a,y_0)+\Phi(\bar{x}+a,y_0)\approx 2\,a\, \partial_x
\Phi(\bar{x},y_0),
\label{forcemi}
\end {eqnarray}
where $\Phi(x,y_0)$ is the local effective interface potential
at the level $y_0$ of the wetting film
(on the top side of the step one has to add $h$ to $y_0$)\/. 
Extending Eq.~\eqref{eqtheta} to inhomogeneous substrates one can
define a spatially varying ``equilibrium contact angle''
$\cos\theq(x) = 1+ \Phi(x,y_0)$\/. In this sense, a droplet in the vicinity of
a topographic step is exposed to an effective chemical wettability
gradient which it follows.
Expanding $\partial_x\Phi(x,y)$ for large $|x|$ yields
\begin{equation}
F_\Pi \approx \mp \frac{3\,a\,C\,h}{\bar{x}^4}+\mathcal{O}(\bar{x}^{-5})
\label{flimit1},
\end{equation}
with the upper sign corresponding to the plus case and the lower
sign to the minus case.
The force is equal on both sides of the step and it increases
linearly with the step height but it decreases rather rapidly with
the distance from the step, however, more slowly than in the case
of nanodroplets. The force increases linearly with the base
width $2\,a$ rather than with the cross-sectional area $A_d$\/. 
As in the case of nanodroplets the actual force on a droplet 
does not diverge in the limit $h\to\infty$\/. 
Using the same approximations as in the case of the step of finite
height, the force on a droplet at a distance $|x|$ from an isolated
edge is given by
%%%In contrast to this, for a step with an infinite step size the disjoining 
%%%pressure is approximated by 
%%%%==========================   EQUATION  ================================ 
%%%%\begin{subequations}
%%%\begin{eqnarray}
%%%\Pi(x,y)= \frac{\pi\Delta M_e}{45\,y^9}
%%%-\frac{H_e}{6\,\pi\,y^3}-\frac{\pi\Delta N_c}{2\,y^4}\\\nonumber
%%%-\frac{\pi\,H_e}{12x^3}+\mathcal{O}(x^{-4})\quad \mbox{for}\quad x\rightarrow\mp\infty\quad
%%%\label{plimit2}
%%%\end{eqnarray}
%%%%\end{subequations}
%%%%=======================================================================
%%%and the net lateral force on the droplet is then given by (see also Ref.~\cite{moosavi08a})
%==========================   EQUATION  ================================ 
\begin{equation}
F_\Pi \sim \mp
\frac{3\,a\,C}{|\bar{x}|^3}+\mathcal{O}(\bar{x}^{-4}),
\label{flimit2}
\end{equation}
with the upper sign corresponding to the plus case and the lower
sign to the minus case. In the vicinity of a wedge the situation is
more complicated. In order to obtain as in Eq.~\eqref{forcemi} 
the effective interface
potential, the point $(x,y)$ corresponding to the upper limit of the
integral there has to correspond to a point at infinite distance from the
substrate. However, taking $y$ to $\infty$ for a fixed value
$\bar{x}$ does not change the distance from the vertical part of
the wedge. At this point it is not clear whether the force integral
in Eq.~\eqref{dpforce} can be approximated by the form given in
Eq.~\eqref{forcemi} because the basic assumption, that the disjoining
pressure is negligible at the apex, is probably not true. 
Expanding the force as calculated from Eq.~\eqref{forcemi} 
for large distances from a step of very large height one obtains
terms of the order $\mathcal{O}(\bar{x}^{-3})$ competing with terms
of order $\mathcal{O}(h\,\bar{x}^{-4})$, which indicates that 
in the case of a wedge the approximations Eq.~\eqref{forcemi} is
based on lead to a mathematically ill-posed problem.

\begin{figure}
\begin{center}
\includegraphics[width=0.5\linewidth]{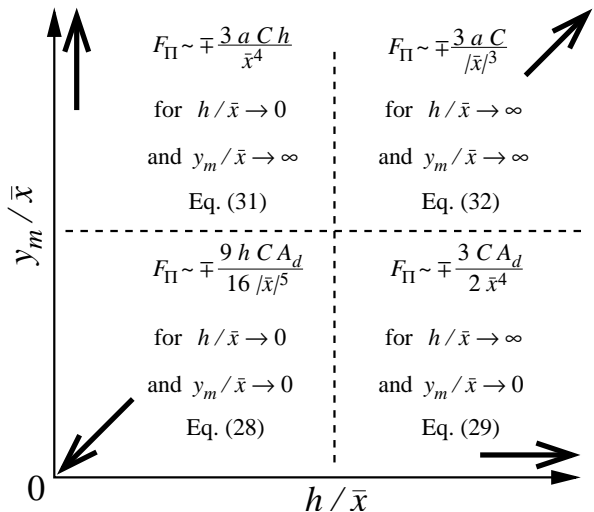}
\end{center}
\caption{\label{asymptofig} The total disjoining pressure induced force
per unit ridge length $F_\Pi=f_\Pi\,A_d$
at large distance from steps depends on two length
ratios $h/\bar{x}$ and $y_m/\bar{x}$ (the cross-sectional area
$A_d$ is proportional to $y_m^2$)\/. This figure summarizes the
analytical results obtained in Subsec.~\ref{direction}\/.}
\end{figure}

In all cases $F_\Pi$ 
essentially depends on the ratio of the
step height $h$ and the distance from the step $\bar{x}$ as well
as on the ratio of the apex height $y_m$ and $\bar{x}$\/. The
asymptotic results are summarized in Fig.~\ref{asymptofig}\/. 
$F_\Pi$ varies according to a power law $\bar{x}^{-\zeta}$,
$\zeta \in \mbox{I}\!\mbox{N}$, of the distance from the step.
For finite sized droplet and steps of finite height ($h/\bar{x}\to 0$
and $y_m/\bar{x}\to 0$) we obtain the fastest decay with
$\zeta=5$\/. For almost macroscopic droplets ($y_m/\bar{x}\to
\infty$) in the vicinity of finite sized steps and for nanodroplets
near isolated edges and wedges one has $\zeta=4$\/. 
For large drops ($y_m/\bar{x}\to \infty$)
next to an isolated edge we get
the weakest decay with $\zeta=3$\/. In any case, the 
total force per unit length $F_\Pi$ is proportional to the Hamaker
constant as observed in the numerical solution of the mesoscopic
Stokes dynamics as well as in the force analysis presented 
in Subec.~\ref{force}\/. 

%%%%as shown in Ref.~\cite{moosavi08a}\/. In all cases (edge, wedge,
%%%%and step) to the given order in $1/x$ the force on a droplet is
%%%%proportional to the Hamaker constant and independent of the
%%%%properties of the coating layer, i.e., only determined by the long
%%%%ranged component of the DJP and not by the equilibrium contact
%%%%angle.

%%%Equations~\eqref{flimit1} and \eqref{flimit2} show that to the 
%%%above order in $1/x$, the force is proportional to the Hamaker constant and
%%%%is diminished. The sign of the force is also depends  on the sign 
%%%%of the Hamaker constant; 
%%%the direction of the motion of the 
%%%nanodroplets is soleley dictated by sign of the Hamaker constant\/. 
%==========================   EQUATION  ================================ 
%=======================================================================
%=========================== SUBSECTION ================================

\subsection{Estimates for the velocity}
The driving force $f$ on the droplets is balanced by viscous 
forces. By applying a simple analysis within the lubrication
approximation one can show that the rate of energy dissipation is
proportional to the square of the velocity $\bar{u}=\partial_t \bar{x}(t)$ 
of the droplets \cite{degennes85,servantie08}\/. 
%Indeed, assuming analyticity of the dissipation rate as a function of
%$\bar{u}$, it has to be $\propto \bar{u}^2$ for small velocities by
%symmetry. 
The form of this dependence can be expected to hold also for
droplets with large contact angles on the basis of analyticity and
symmetry arguments.  By equating this dissipation with the work
$\bar{u}\,f\,\Omega_d$ done by the driving force one finds
\begin{equation}
\bar{u}=d\bar{x}/dt \sim f/\Omega_d.
\end{equation}
For droplets far from the step with $f(\bar{x})$ given by  Eqs.~\eqref{flimit1} 
and \eqref{flimit2} as a power law one has $\bar{x}(t)^{\nu}\sim t$
and therefore
%==========================   EQUATION  ================================ 
\begin{equation}
|\bar{x}(t)|=(|x_0|^{\nu}+c\,t)^{1/\nu}
\label{xt}
\end{equation}
%=======================================================================
with $\nu=4$ for large droplets in the vicinity of edges, $\nu=5$
for large droplets  in the vicinity of steps of finite
height as well as for nanodroplets near isolated edges and wedges,
and $\nu=6$ for nanodroplets near steps of finite height. $c$ is a
constant which also depends on whether there is an edge, wedge, or
a step and whether the droplet is large or small.
%==========================   EQUATION  ================================ 
%%%Based on Eq.~\eqref{xt} the lateral positions of the nanodroplets 
%%%far from the step should follow a power law curve. 
The functional form given by Eq.~\eqref{xt} with the corresponding value
of $\nu$ can be fitted to the positions of
nanodroplets as a function of time obtained by numerically
solving the mesoscopic hydrodynamic equations, e.g., to the data
shown in Fig.~\ref{effstepplus} (droplet on the top side of steps, plus
case) and Fig.~\ref{wedgediffhminus} (droplet on the step base,
minus case)\/. However, the numerically available range of
$\bar{x}$ values is rather
small so that the fits are consistent with the above values of
$\nu$ but cannot rule out different ones.
In addition, it is not clear whether the distances
considered in the numerical solutions of the mesoscopic
hydrodynamic equations are large enough to reach the asymptotic
regime considered here, and whether the droplets should be
considered small or large in the above sense.

\section{Perspectives}

A major and obvious driving force for studying the dynamics of nanodroplets on
structured substrates is the rapid development and miniaturization
of microfluidic devices, in particular of open microfluidic devices
\cite{zhao01,lam02,zhao02,zhao03}\/. But this is not the only
research area
for which a detailed understanding of the influence of the
long-ranged part of the intermolecular interactions on fluids in
the vicinity of lateral surface structures might be important.
Another example is the dynamics of nanodroplets at chemical surface
structures as discussed in Refs.~\cite{moosavi08a,moosavi08b}\/.
Moreover, dewetting processes are also strongly influenced by surface
heterogeneities, both during the initial phase of film breakup
\cite{kargupta02c,mukherjee07} as well as during hole
growth \cite{ondarcuhu05}\/. The latter example is particularly
interesting in this respect because it reveals an intrinsic
nanoscopic length scale which has to be understood: the receding
contact line is pinned only by steps of a minimum height which
increases with the size of the liquid molecules \cite{ondarcuhu05}---a clear
indication that details of the intermolecular interactions in the
vicinity of the step are relevant. 

All results presented in this article have been obtained for 
homogeneous straight
liquid ridges. Apart from the fact that such ridges are unstable
with respect to breaking up into three-dimensional droplets
\cite{davis80,brinkmann05,koplik06a,mechkov08}, the question
remains to check how relevant
these results are for actual three-dimensional droplets.
In this context we point out that the basic driving mechanism for
droplets in the vicinity of steps is the difference of the
disjoining pressure on that side of the droplet which is closer to
the step and the side which is further away from the step. In such
a situation also three-dimensional droplets move. However, the
third dimension 
certainly changes the behavior of droplets spanning topographic
steps \cite{brinkmann04b,herminghaus08}: depending on the droplet volume, the
droplet can spread along the step into a cigar shaped
configuration. The influence of the long-ranged part of the
intermolecular interactions on this phenomenon has not yet been studied. 

It is worthwhile to point out that, although the dynamics is
different, there are strong similarities between nanodroplets and
solid nanoclusters: their energetics on
structured surfaces is determined by intermolecular forces as
demonstrated in Ref.~\cite{yoon03} by molecular dynamics simulations
of gold clusters on graphite surfaces. Unfortunately, in such
simulations taking into account the long-ranged component of the
intermolecular forces increases the numerical cost drastically, such
that most of the effects discussed here are not accessible  by
molecular dynamics simulations \cite{deconinck08}\/.

This leaves the question of experimental tests of our theoretical
predictions presented here. 
As detailed in Ref.~\cite{moosavi06b} the forces on the
nanodroplets are of the order of $10^{-13}$~N (i.e., about eight
orders of magnitude stronger than the gravitational force on such a
droplet) and the resulting velocities range between
$0.1\,\mbox{mm}/\mbox{s}$ and
$0.1\,\mu\mbox{m}/\mbox{s}$ for viscosities between
$0.1\,\mbox{Pa}\,\mbox{s}$ and $100\,\mbox{Pa}\,\mbox{s}$\/.
While topographic surface structures of
almost any type can be produced with modern lithographic
techniques,
positioning nanodroplets with nanometer accuracy next to a step
remains a tough challenge. Most promising are techniques based on
using atomic force microscopes as pens \cite{yang06,fang06}, but
there are no experiments available yet. Experimentally it is
much easier to grow droplets from an aerosol or a vapor phase
rather than to deposit them at a specific location. 
Experimentally it has been
shown, that water nanodroplets preferentially condense onto terrace
steps of
vicinal surfaces \cite{hu96}\/. However, these data do not allow
one to
determine whether the droplets reside on the top terrace, on the
bottom terrace, or whether they span the step. This example
shows, that condensation on (and evaporation from) nano-structured
substrates is a challenging problem of its own.

\ack
{M. Rauscher acknowledges financial support by the Deutsche
Forschungsgemeinschaft (DFG) within the priority program SPP 1164 "Nano- and
Microfluidics".}

\appendix

\section{Numerical algorithm }
\label{numersec}
%==========================   FIGURE  ================================== 
\begin{figure}
\includegraphics[width=0.8\linewidth]{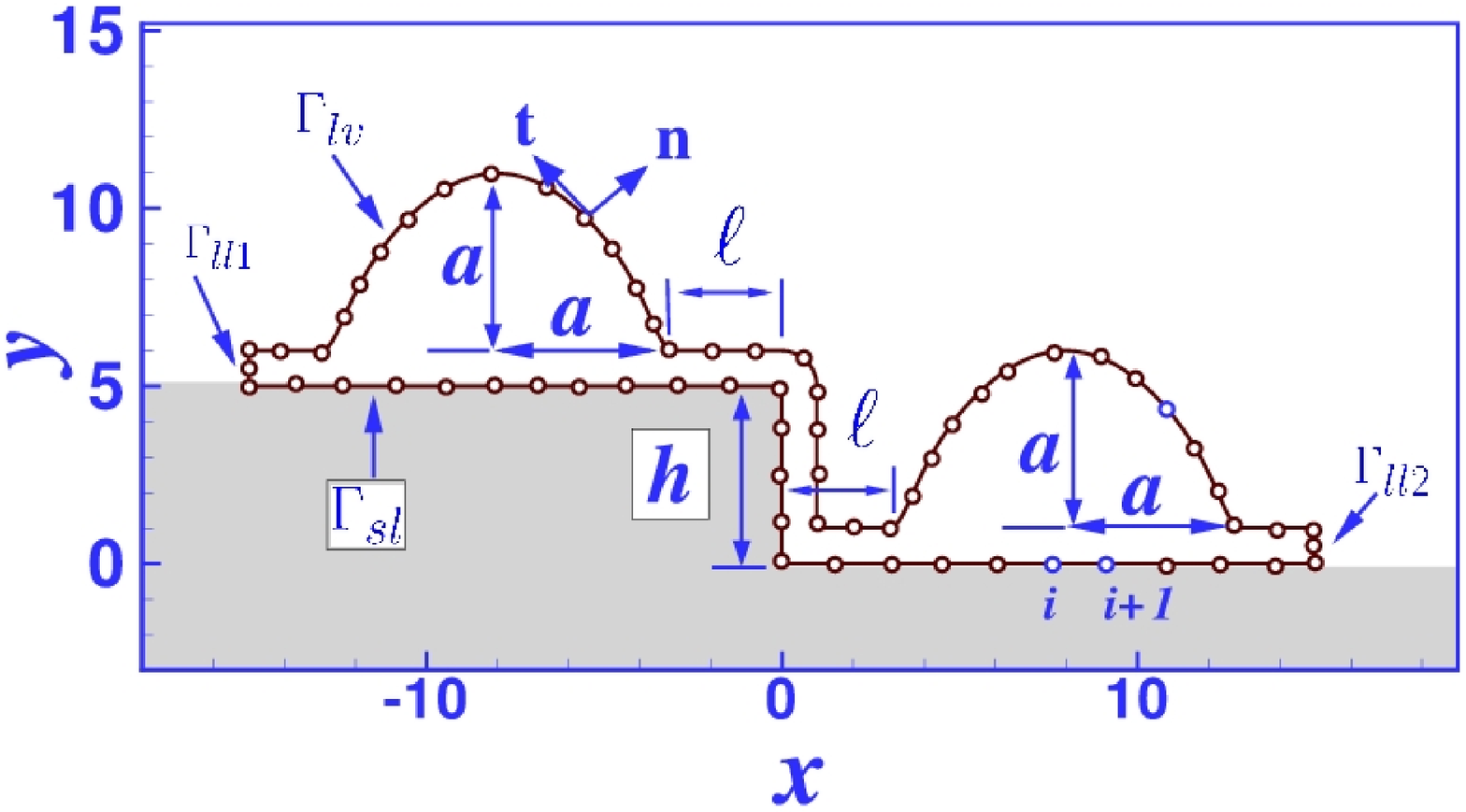}
\caption{The dynamics of a droplet is investigated which initially
is positioned either on the top side or the bottom side of a step
at a distance $\ell$ from the step.
The boundary of the system $\Gamma$ decomposes
into three different groups: liquid-liquid 
($ll$), liquid-solid ($ls$), and liquid-vapor ($lv$) interfaces. 
The discretized node points considered in the numerical investigation
are indicated; $\vct{n}$ and $\vct{t}$ represents the 
 normal and the tangential unit vectors on $\Gamma$, respectively.} 
%$\hat{p}$ and  $\hat{q}$ are two arbitrary points on the boundary of the system.}
\label{numfig}
\end{figure}
In order to study the effect of the intermolecular 
forces on nanodroplets near a topographic step
we solve Eqs.~\eqref{eq:stokes}--\eqref{eq:surfacebc} numerically
using a standard and accurate biharmonic 
boundary integral method (BBIM)
\cite{kelmanson83a,betelu97,mazouchi04,moosavi06b,moosavi08a,moosavi08b}\/.
%%%%==========================   EQUATION  ================================ 
%%%\begin{equation}
%%%\label{continuity}
%%%\nabla\cdot\textbf{u}=0
%%%\end{equation} 
%%%%=======================================================================
%%%%==========================   EQUATION  ================================ 
%%%\begin{equation}
%%%\label{stokes}
%%%\nabla^2{\textbf{u}}=\frac{1}{\textnormal{C}}\nabla\,p\,,
%%%\end{equation}
%%%%=======================================================================
%%%where ${\bf{u}}=(u_{x},u_{y})$ is the velocity vector and  $p$ 
%%%represents the pressure. Note that, since we have considered the disjoining pressure as an imposed normal stress  \cite{moosavi06b,oron97},the disjoining pressure does not enter in Eq.~\eqref {stokes} and only appear in the boundary condition (see the below). However, one may equivalently consider the disjoining pressure as 
%%%a body force \cite{moosavi06b,oron97}. 
%%%To convert the equations 
%%%to non-dimensional form velocity has been scaled with $Ab/\mu$, 
%%%where $\mu$ is the viscosity. 
%%%Length and pressure have also been 
%%%scaled with $b$ and $\gamma/b$, respectively, as already discussed. 
To this end we introduce the stream function $\psi(x,y)$ so that
$\partial{\psi}/\partial{y}=u_{x}$ and 
$\partial{\psi}/\partial{x}=-u_{y}$ as well as the vorticity 
$\omega(x,y)={\partial{u_x}/\partial
{y}}-{\partial{u_y}}/{\partial{x}}$ which allows us to reformulate
the dimensionless
versions of Eqs.~\eqref{eq:stokes} and \eqref{eq:incomp}  
in terms the following  harmonic and 
biharmonic equations \cite{kelmanson83a,betelu97,mazouchi04,moosavi06b,moosavi08a,moosavi08b}:
%==========================   EQUATION  ================================ 
\begin{equation}
\label{omega}
 \nabla^{2}\omega=0
\end{equation} 
%=======================================================================
and
%==========================   EQUATION  ================================ 
\begin{equation}
\label{psi}
\nabla^4 \psi=0\,\cdot
\end{equation} 
%=======================================================================
%or equivalently, from the above set of equations, one can show that $\nabla^4\psi=0$. 
%As will be discussed later (see the appendix) Eqs.~\eqref{omega} 
%and \eqref{psi} are more appropriate for our numerical algorithm 
%than the usual continuity and Stokes equations. However, 
%\cite{moosavi06b,oron97}
%As a consequence of the mass conservation and
%incompressibility the velocity of the interface is given by the normal component of the flow field at the interface.

The standard BBIM relies on mapping the equations for $\omega$ and
$\psi$ onto the boundary $\vct{r}(s)=(x(s),y(s))$ of the fluid,
parameterized in terms of its contour length parameter $s$\/.  This
results in an integral
equation for $\omega$, $\psi$, and their derivatives
$\omega_n=\vct{n}\cdot\grad\omega$ and
$\psi_n=\vct{n}\cdot\grad\psi$, with the surface normal vector
pointing outwards of the liquid. By dividing the  boundary of 
the system into a series of elements (see Fig.~\ref{numfig}) one
obtains a coupled system of algebraic equations which can be solved
numerically. With the tangential velocity $u_t=\psi_n$ and the 
normal velocity $u_n=-\psi_s$ (with the index $s$ indicating
the derivative in the direction tangential to the boundary) 
the position of the liquid boundary after a time step can be calculated 
via the explicit Euler scheme
\cite{kelmanson83a,betelu97,mazouchi04,moosavi06b,moosavi08a,moosavi08b}:
%==========================   EQUATION  ================================ 
\begin{equation}
\textbf{r}(t+\Delta t)=\textbf{r}(t)+\textbf{u(t)}\Delta t.
\end{equation} 
%=======================================================================

In order to solve these equations the boundary conditions of the system 
must be expressed in terms of $\omega$ and $\psi$\/. Depending on 
the phases in contact with each other, three different types of 
boundary interfaces can be identified (see Fig.~\ref{numfig}): 
liquid-solid interfaces $\Gamma_{ls}$, liquid-liquid interfaces (those 
boundaries $\Gamma_{ll1}$ and $\Gamma_{ll2}$ which are located at
the end sides of the system), and liquid-vapor interfaces
$\Gamma_{lv}$\/. For $\Gamma_{ls}$ we impose the no-slip
condition ($\vct{u}=0$) which corresponds to $\psi=0$ 
and $\psi_n=0$\/. For $\Gamma_{ll1}$ and $\Gamma_{ll2}$ 
we apply a no-flux condition which corresponds to having a vertical
symmetry plane there. Such a system corresponds to a periodic repetition
of the system attached to its mirror image. Correspondingly the 
slope of the liquid-vapor interface at the side ends of the system is zero. 
These conditions can be implemented by setting $\psi=0$ and
$\omega=0$ there.
The tangential and the normal component of the 
boundary condition \eqref{eq:surfacebc}
along the liquid-vapor interface $\Gamma_{lv}$ 
in terms of the stream function and the vorticity read (lower
indices $s$ indicate derivatives  with respect to the contour
length parameter $s$)
%%%there is no 
%%%specific value for the unknown variables ($\psi$, $\psi_n$, $\omega$, 
%%%and $\omega_n$) but two of these variables can be interpreted in 
%%%terms of the other two variables using the fact that the shear 
%%%stress on the interface is zero and the normal stresses are 
%%%balanced by  surface tension and the disjoining pressure. 
%%%In non-dimensional form one has
%%%%==========================   EQUATION  ================================ 
%%%\begin{equation}
%%%\sum_{i,j} \sigma_{ij}n_{i}t_{j}=0
%%%\label{sb}
%%%\end{equation}
%%%%=======================================================================
%%%%==========================   EQUATION  ================================ 
%%%\begin{equation}
%%%\sum_{i,j}\sigma_{ij}n_{i}n_{j}=-\kappa+\Pi\,-f\,x,
%%%\label{nb}
%%%\end{equation}
%%%where $\sigma_{ij}$ is the stress tensor $\sigma_{ij}=-p\delta_{ij}+
%%%C\,(\partial u_i/\partial x_j+\partial u_j/\partial x_i)$  and $f$ describes 
%%%an external body force in the direction $x$ (per unit volume: in units of $\sigma/b^2$). 
%%%$\textbf{t}$ and $\textbf{n}$ represent the tangential and 
%%%normal unit vectors of the interface, respectively. $\kappa$ 
%%%stands for the local curvature and can be calculated from  
%%%%==========================   EQUATION  ================================ 
%%%Expressing Eqs.~\eqref{sb} and \eqref{nb} in terms of $\omega$ and $\psi$ one has 
\begin{equation}
\omega=2\,\psi_{ss}+2\,\kappa\,\psi_{n}
\end{equation}
and
\begin{equation}
\omega_{n}=-2\,{\psi_n}_{ss}+2\,\kappa\,\psi_{ss}+2\,\kappa_{s}\,\psi_{s}
+\frac{\kappa_{s}+\Pi_s+g\,x_s}{C}\,,
\end{equation}  
%=======================================================================
respectively, with the local curvature 
\begin{equation}
\kappa=-\frac{y_{ss}\,x_s-x_{ss}\,y_s}{(x_s^2+y_s^2)^{3/2}}.
\end{equation} 

In order to increase the efficiency of the numerical calculations
we employ an adaptive time stepping: for any numerical step, 
the time step is selected such that the displacement of any node does 
not exceed $\delta$ percents of the length of the 
elements connected to that node; $\delta$ can be changed during the 
numerical calculations. 
The starting value for $\delta$ and the rate of its increase depends 
on the actual situation but typically we have started with $\delta=0.01$ and 
then gradually increased it to 0.1 or even more. 
In order to avoid numerical instabilities, the position of 
the end points of the boundary elements are smoothed after a specified
number of steps by fitting
a spline through the points on the liquid-vapor interface, followed
by selecting new and equally spaced points on the spline. 

%%%
%%%shape relaxation of the droplets (which can lead to significant lateral displacements \cite{ondarcuhu92}) the initial droplet shape 
%%%was also chosen to be close to a parabola
%%%but connecting smoothly to the precursor film:
%%%%==========================   EQUATION  ================================ 
%%%\begin{equation}
%%%\label{inicond}
%%%y(x;t=0)= y_0+a\,\left[1-\left(\frac{|x-\bar{x}|}{a}\right)^2\,
%%%\right]^{|x-\bar{x}|^m+1},
%%%\end{equation} 
%%%with $a$ as the droplet height at the center and half the base width. 
%%%The distance of the droplet edge from the step is then given by 
%%%$\ell=|\bar{x}|-a$  with $|\bar{x}|$ the position of the center 
%%%of the droplet in the $x$-direction (see Fig.~\ref{edgediffb})\/. 
%%%The parameter $m$ specifies the smoothness of the transition region 
%%%from the drop to the wetting layer. In this study we choose $m$ to be $10$\/. 
%%%
%%%The dynamics of the droplets was investigated in two different scenarios. 
%%%In the first scenario the droplets were positioned on top of the steps with their 
%%%three-phase contact line $(x=\bar{x}+a,\,y=h+y_0,z)$ at a distance 
%%%$\ell=-\bar{x}-a$ from the steps and in the second scenario the 
%%%droplets were positioned at the base of the steps with their three-phase contact 
%%%line $(x=\bar{x}-a,\,y=y_0,z)$ at a distance $\ell=\bar{x}-a$ from the steps. 

%\bibliography{references}
%\bibliographystyle{my_jpcm}

\end{document}